\documentclass{aa}	
\usepackage[varg]{txfonts}

\usepackage{natbib}
\bibpunct{(}{)}{;}{a}{}{,} 

\usepackage{array}
\usepackage{graphicx}
\usepackage{multirow}

\newcommand\MyBox[2]{
	\fbox{\lower0.75cm
		\vbox to 1.7cm{\vfil
		\hbox to 1.7cm{\hfil\parbox{1.4cm}{#1\\#2}\hfil}
		\vfil}%
	}%
}

\title{Identifying Galaxy Mergers in Observations and Simulations with Deep Learning}
\author{W.~J.~Pearson\inst{\ref{inst:SRON}, \ref{inst:Kapteyn}},
L.~Wang\inst{\ref{inst:SRON}, \ref{inst:Kapteyn}},
J.~W.~Trayford\inst{\ref{inst:Leiden}},
C.~E.~Petrillo\inst{\ref{inst:Kapteyn}},
F.~F.~S.~van~der~Tak\inst{\ref{inst:SRON}, \ref{inst:Kapteyn}}
}
\institute{SRON Netherlands Institute for Space Research, Landleven 12, 9747 AD, Groningen, The Netherlands\label{inst:SRON}\\\email{w.j.pearson@sron.nl}
\and Kapteyn Astronomical Institute, University of Groningen, Postbus 800, 9700 AV Groningen, The Netherlands\label{inst:Kapteyn}
\and Leiden Observatory, Leiden University, P.O. Box 9513, 2300 RA Leiden, The Netherlands\label{inst:Leiden}
}
\authorrunning{W.~J.~Pearson et al.}

\date{Received DD Month YYYY /
Accepted DD Month YYYY}
\abstract{Mergers are an important aspect of galaxy formation and evolution. With large upcoming surveys, such as \textit{Euclid} and LSST, accurate techniques that are fast and efficient are needed to identify galaxy mergers for further study.}
{We aim to test whether deep learning techniques can be used to reproduce visual classification of observations, physical classification of simulations and highlight any differences between these two classifications. With one of the main difficulties of merger studies being the lack of a truth sample, we can use our method to test biases in visually identified merger catalogues.}
{A convolutional neural network architecture was developed and trained in two ways: one with observations from SDSS and one with simulated galaxies from EAGLE, processed to mimic the SDSS observations. The SDSS images were also classified by the simulation trained network and the EAGLE images classified by the observation trained network.}
{The observationally trained network achieves an accuracy of 91.5\% while the simulation trained network achieves 65.2\% on the visually classified SDSS and physically classified EAGLE images respectively. Classifying the SDSS images with the simulation trained network was less successful, only achieving an accuracy of 64.6\%, while classifying the EAGLE images with the observation network was very poor, achieving an accuracy of only 53.0\% with preferential assignment to the non-merger classification. This suggests that most of the simulated mergers do not have conspicuous merger features and visually identified merger catalogues from observations are incomplete and biased towards certain merger types.}
{The networks trained and tested with the same data perform the best, with observations performing better than simulations, a result of the observational sample being biased towards conspicuous mergers. Classifying SDSS observations with the simulation trained network has proven to work, providing tantalizing prospects for using simulation trained networks for galaxy identification in large surveys.}


\begin{document}
\maketitle

\section{Introduction}
Galaxy-galaxy mergers are of fundamental importance to our current understanding of how galaxies form and evolve in cold dark matter cosmology \citep[e.g.][]{2014ARA&A..52..291C}. Dark matter halos and their baryonic counterparts merge under hierarchical growth to form the universe that we see today \citep{2015ARA&A..53...51S}. Mergers play an important role in many aspects of galaxy evolution such as galaxy mass assembly, morphological transformation and growth of the central black hole \citep[e.g.][]{1996ApJ...465..278J, 2003ApJ...597..893N, 2006ApJS..163....1H, 2008ApJ...680..295B, 2008MNRAS.384....2G, 2009ApJ...701.2002G}. In addition, galaxy mergers are believed to be the triggering mechanism of some of the brightest infrared objects known: (ultra) luminous infrared galaxies \citep{1996ARA&A..34..749S}. With bright infrared emission often comes high star formation rates (SFRs), hence a prevailing interpretation from early merger works is that most mergers go through a starburst phase \citep[e.g.][]{1985MNRAS.214...87J, 2005ASSL..329..143S}.

Recent studies have begun to dismantle the claim that all galaxy mergers are starbursts. In a study of 1500 galaxies, within 45~Mpc of our own, \citet{2015MNRAS.454.1742K} have found that the increase in SFR in merging galaxies is at most a factor of two, with the majority of galaxies showing no evidence of an increase in SFR, or even showing evidence of mergers quenching the star formation. Galaxy mergers do still cause starbursts and a higher fraction of starbursts are mergers than starbursting non-mergers \citep{2014ApJ...789L..16L, 2015ApJ...807L..16K, 2017A&A...607A..70C}. Claims about the importance of mergers depend critically on our ability to recognise galaxy interactions. A method to reliably identify complete merger samples among a large number of galaxies is clearly needed.

Existing automated techniques for detecting mergers include selecting close galaxy pairs or selecting morphologically disturbed galaxies. The close pair method finds pairs of galaxies that are close, both on the sky and in redshift \citep[e.g.][]{2000ApJ...530..660B, 2002ApJ...565..208P, 2003MNRAS.346.1189L, 2004ApJ...617L...9L, 2005AJ....130.1516D}. This method requires highly complete, spectroscopic observations and, as a result, is observationally expensive. It can also be contaminated by flybys \citep{2012ApJ...751...17S, 2014ApJ...790L..33L}. Selecting the morphologically disturbed galaxies using quantitative measurements of non-parametric morphological statistics, such as the Gini coefficient, the second-order moment of the brightest 20 percent of the light \citep{2004AJ....128..163L} and the CAS system \citep[e.g.][]{2000AJ....119.2645B, 2000ApJ...529..886C, 2001defi.conf..170W, 2003AJ....126.1183C}, aims to detect disturbances such as strong asymmetries, double nuclei or tidal tails. This method relies on high-quality, high-resolution imaging to detect these features beyond the local universe and has a high percentage of misclassifications (> 20\%), especially at high redshift \citep{2015ApJS..221....8H}. There is also the option to classify galaxies through visual inspection. However, visual classifications are hard to reproduce and are time consuming. Large crowd sourced methods, such as Galaxy Zoo\footnote{http://www.galaxyzoo.org/} \citep{2008MNRAS.389.1179L}, are not scalable to the sizes of the data sets expected from upcoming surveys. Visual identification can also suffer from low accuracy and incompleteness \citep{2015ApJS..221....8H}.

Deep learning techniques have the potential to revolutionise galaxy classification. Once properly trained, the neural networks used in deep learning can classify thousands of galaxies in a fraction of the time it would take a human, or team of humans, to classify the same objects. The use of deep learning for galaxy classification was brought to wider attention after Galaxy Zoo lead a competition on the Kaggle platform\footnote{https://www.kaggle.com/c/galaxy-zoo-the-galaxy-challenge}, known as The Galaxy Challenge, to develop a machine learning algorithm to replicate the human classification of the Sloan Digital Sky Survey \citep[SDSS;][]{2000AJ....120.1579Y} images. This competition was won by \citet{2015MNRAS.450.1441D} using a deep neural network, the architecture of which has formed the base for subsequent deep learning algorithms for galaxy classification \citep[e.g.][]{2015ApJS..221....8H, 2017MNRAS.472.1129P}. More recently, deep learning has been applied to SDSS images from Galaxy Zoo to classify objects as merging or non-merging systems using transfer learning, that is taking a pre-trained network and retraining the output layer to classify images into a different set of classifications \citep{2018MNRAS.479..415A}. There has also been work using deep learning techniques to identifying mergers and tidal features in optical data from the Canada-France-Hawaii Telescope Legacy Survey \citep{2012AJ....143...38G, 2019MNRAS.483.2968W}. These techniques will have an important use in classifying galaxies in large, upcoming surveys, such as the Large Synoptic Survey Telescope \citep[LSST;][]{2009arXiv0912.0201L} or \textit{Euclid} \citep{2011arXiv1110.3193L}.

In this work, we aim to develop a neural network architecture and independently train it with two different training sets. This will result in a trained neural network that can identify visually classified mergers from the SDSS data as well as one that can identify physically classified mergers from the Evolution and Assembly of GaLaxies and their Environments (EAGLE) hydrodynamical cosmological simulation \citep{2015MNRAS.446..521S}. Once trained, the networks will be cross applied: SDSS images through the EAGLE trained network and images of simulated galaxies from EAGLE through the SDSS trained network. Visually identified merger catalogues constructed from surveys, such as the SDSS, are biased towards mergers that produce conspicuous features but cosmological simulations include a wide variety of merging galaxies with different mass ratios, gas fractions, environments, orbital parameters etc. Therefore, through training our neural network separately with visual classifications of real observations, physical classifications in simulations and the cross-applications of the two, we can better understand any potential biases in observations and identify problems in simulations in terms of reproducing realistic merger properties.

The paper is structured as follows: Sect. \ref{sec:data} describes the data sets used, Sect. \ref{sec:cnn} covers the neural networks, Sect. \ref{sec:results} provides the results and discussion and Sect. \ref{sec:conclusion} the concluding remarks. Where necessary, Wilkinson Microwave Anisotropy Probe year 7 (WMAP7) cosmology \citep{2011ApJS..192...18K, 2011ApJS..192...16L} is followed, with $\Omega_{M}$ = 0.272, $\Omega_{\Lambda}$ = 0.728 and H$_{0}$ = 70.4~km~s$^{-1}$~Mpc$^{-1}$.

\section{Image Data}\label{sec:data}
\subsection{SDSS Images}\label{subsec:sdss}
To train the neural network, a large number of images of merging and non-merging systems are required. For the training the observational network, we create our merger and non-merger samples by following \citet{2018MNRAS.479..415A} and combining the \citet{2010MNRAS.401.1552D, 2010MNRAS.401.1043D} merger catalogue with non-merging systems. The \citet{2010MNRAS.401.1552D, 2010MNRAS.401.1043D} catalogue contains 3003 merging systems selected by visually rechecking the visual classifications of all objects from Galaxy Zoo with the fraction of people who classified the object as merging greater than 0.4 and spectroscopic redshifts between 0.005 and 0.1. As a result of this thorough visual classification, the \citet{2010MNRAS.401.1552D, 2010MNRAS.401.1043D} catalogue is likely to be conservative and mainly contain galaxies with obvious signs of merger, i.e. two (or more) clearly interacting galaxies or obviously morphologically disturbed systems, and may miss more subtle mergers. The SDSS spectra were only taken for objects with apparent magnitude r~$<17.77$, or absolute magnitude r~$<-20.55$ at $z=0.1$ hence resulting in an effective mass limit of $\approx 10^{10}$~M$_{\odot}$ at $z=0.1$ \citep{2010MNRAS.401.1043D}. For the non-merging systems, we generated a catalogue of all SDSS objects with spectroscopic redshifts in the same range as the \citet{2010MNRAS.401.1552D, 2010MNRAS.401.1043D} catalogue and the fraction of people who classified the object as merging in Galaxy Zoo less than 0.2 and then randomly selected 3003 of these to form the sample. As we also require spectroscopic redshifts, the non-merger sample will have the same effective mass limit of $\approx10^{10}$~M$_{\odot}$.

\begin{figure}
	\centering
	\includegraphics[width=0.5\textwidth]{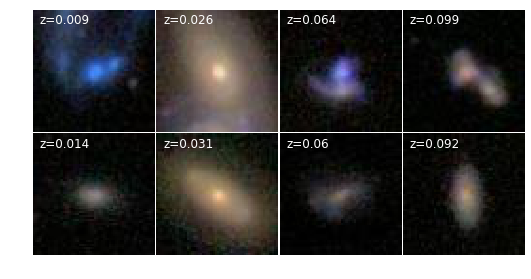}
	\caption{Examples of the central 64$\times$64 pixels of SDSS gri, as blue, green and red respectively, galaxy images, corresponding to an angular size of 25.3$\times$25.3 arcsec. The top row shows merging galaxies from the \citet{2010MNRAS.401.1552D, 2010MNRAS.401.1043D} catalogue while the bottom row shows non-merging galaxies.}
	\label{fig:sdssimg}
\end{figure}

Cut-outs of the merging and non-merging objects were then requested from the SDSS jpeg cut-out server for data release 7\footnote{\url{http://cas.sdss.org/dr7/en/tools/chart/default.asp}} (DR7) to create 6006 images with the gri bands as the blue, green and red colour channels respectively, each of 256$\times$256 pixels. SDSS images created this way use a modified version of the \citet{2004PASP..116..133L} non-linear colour normalisation. These images were then cropped to the centre 64$\times$64 pixels for use to reduce memory requirements while training. Larger image sizes were tested but showed no clear improvements over 64$\times$64 pixel images. Examples of the central 64$\times$64 pixels of merging and non-merging SDSS galaxies are given in Fig. \ref{fig:sdssimg}.

The SFR and M$_{\star}$ for the SDSS objects were gathered from the MPA-JHU catalogue\footnote{\url{https://wwwmpa.mpa-garching.mpg.de/SDSS/DR7/}}; the M$_{\star}$ were created following the techniques of \citet{2003MNRAS.341...33K} and \citet{2007ApJS..173..267S}, while the SFR were based on the \citet{2004MNRAS.351.1151B} catalogue. The redshifts and ugriz magnitudes come from the SDSS DR7. \citet{2010MNRAS.401.1043D} find that 54\% are major mergers, defined as the ratio of the masses of the merging galaxies is between 1/3 and 3.

\subsection{EAGLE Images}\label{subsec:eagle}
For the simulation network, simulated gri images from EAGLE were used \citep{2015MNRAS.450.1937C, 2016A&C....15...72M}. These single channel images were generated using the \texttt{py-SPHviewer} code \citep{2017ascl.soft12003B}. EAGLE star particles are treated as individual simple stellar populations in the imaging, assuming a \citet{2003PASP..115..763C} initial mass function. The luminosity of particles in each band is obtained using the \citet{2003MNRAS.344.1000B} population synthesis model via their zero-age mass, current age and smoothed particle hydrodynamics metallicity. Particle smoothing lengths are calculated based on the 64th nearest neighbour, as described in \citet{2017MNRAS.470..771T}. We note that attenuation by dust is not accounted for, with these primary images functioning as maps of pure stellar emissivity.

To provide enough galaxies to adequately train a neural network, EAGLE galaxies from the simulation snapshots with a redshift of less than 1.0 were used. Objects with stellar mass (M$_{\star}$) greater than $10^{10}$~M$_{\odot}$ were selected while the merging partner of the merging systems must be larger than $10^{9}$~M$_{\odot}$. The merging partner must also be more than 10\% of the M$_{\star}$ of the primary galaxy. Galaxies were deemed to have merged when they are tracked as two galaxies in one simulation snapshot and then tracked as one galaxy in the following snapshot in the EAGLE merger trees catalogue \citep{2017MNRAS.464.1659Q}. This prevents the inclusion of chance flybys that may be selected as mergers if the EAGLE galaxies were selected based on proximity. Systems that are projected to merge, using a closing velocity extrapolation, within the next 0.3~Gyr (pre-merger) or are projected to have merged, again using a closing velocity extrapolation of the progenitors, within the last 0.25~Gyr (post-merger) were selected, along with a number of non-merging systems, and gri band images were created of these systems. \citet{2005MNRAS.361..776S} have shown that the effects of a merger are visible for approximately 0.25~Gyr after the merger event while the pre-merger stage is much longer. However, we chose to have the pre and post merger period approximately equal as tests conducted with longer pre-merger times showed no improvement, see discussion in Sect. \ref{sec:sim_train_net}. We note, however, that the merger timing may suffer from imprecision as a result of the coarse time resolution of the EAGLE simulation, i.e. the time between snapshots, which becomes coarser at lower simulation redshift. Each galaxy is imaged at an assumed distance of 10~Mpc and each image contains all material within 100~kpc of the centre of the target galaxy and is 256$\times$256 pixels, where 256 pixels corresponds to a physical size of 60~kpc. There are 537 pre-merger, 339 post-merger and 335 non-merging systems, each with six random projections to increase the size of the training set. Each of the six projections are treated as individual galaxies resulting in 3222 pre-merger, 2034 post-merger and 2010 non-merging galaxy images for training. The pre-mergers and post-mergers were combined to form the merger class, keeping the pre-merger image if the same galaxy appears in both sets.

To make the raw EAGLE images look like SDSS images (processed EAGLE images), a number of operations were performed. For each projection of each system, a redshift was randomly chosen from the redshifts of the objects in the \citet{2010MNRAS.401.1552D, 2010MNRAS.401.1043D} catalogue and the surface brightness of the galaxy was corrected to match this redshift. The image was also re-binned using interpolation with the python \texttt{reproject} package \citep{astropy-reproject} so that the physical resolution of the EAGLE image matches that physical resolution of an SDSS galaxy at the selected redshift. The resulting apparent r-band magnitudes are less than 17.77 for all but 58 of the 10\,134 galaxy projections, meaning that the brightness of the simulated galaxies is consistent with the observed SDSS galaxies. Once the surface brightness and physical resolution correction was completed, the observed SDSS point spread functions (PSFs) for the gri bands were created using the stand alone PSF tool\footnote{\url{https://www.sdss.org/dr12/algorithms/read_psf/}} and the simulated images were convolved with these PSFs. The EAGLE galaxies were then injected into real SDSS images to add realistic noise. Finally, red, green, blue (RGB) images were generated from the gri bands using a modified \citet{2004PASP..116..133L} non-linear colour normalisation to closely match the way the SDSS RGB images are made. A brief comparison using a simple linear colour scaling to generate the EAGLE images is discussed in Appendix \ref{app:linear}.

\begin{figure}
	\centering
	\includegraphics[width=0.5\textwidth]{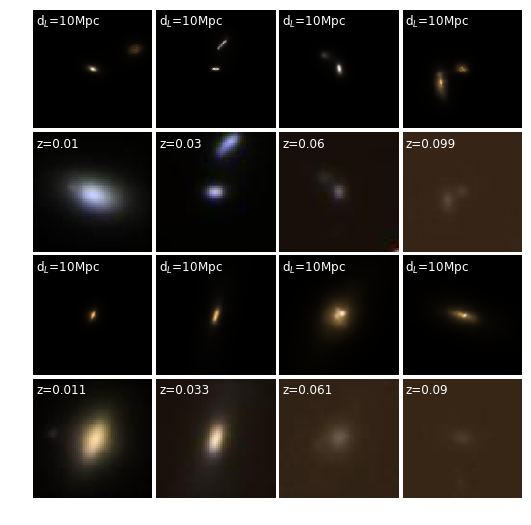}
	\caption{Examples of the raw (first and third rows, linear colour scaling) and processed (second and fourth rows, \citet{2004PASP..116..133L} non-linear colour scaling) EAGLE images for merging (first and second rows) and non-merging (third and fourth rows) systems. The raw images shown are 128$\times$128 pixels and imaged at 10~Mpc, corresponding to a \textit{physical} size of 30$\times$30 kpc or an \textit{angluar} size of 621$\times$621 arcsec, while the processed images are 64$\times$64 pixel images corresponding to an \textit{angular} size of 25.3$\times$25.3 arcsec. The redshifts are those that the EAGLE images have been projected to.}
	\label{fig:eagleimg}
\end{figure}

To get real SDSS noise, the position of all known SDSS objects from DR7 in three SDSS images were collected. The noise images were generated by offsetting from the position of the objects in these images by a random distance between 6.329 and 18.986 arcsec (i.e. between 0.25 and 0.75 times the average separation of SDSS objects) and with a random angle in the RA-dec plane. Then 256$\times$256 pixel cut-outs were made, centred on these offset positions, and were used as noise in the EAGLE images. The code used to make the EAGLE images SDSS like and get the noise cut-outs can be downloaded from GitHub\footnote{\label{ft:git}\url{https://github.com/wjpearson/SDSS-EAGLE-mergers}} while examples of the raw and processed EAGLE images can be found in Fig. \ref{fig:eagleimg}.

The M$_{\star}$, star formation rate (SFR), ugriz absolute magnitudes, galaxy asymmetry, merger mass ratio and time to or since the merger event of the EAGLE galaxies are from the simulation. For the merging systems, M$_{\star}$ and SFR are calculated for the merger remnant. The galaxy asymmetry is the 3D asymmetry and is calculated as described in \citet{2019MNRAS.483..744T}. Uniform bins of solid angle were created about the galaxy centre and the M$_{\star}$ within each bin is summed. The asymmetry is then the sum of the absolute mass difference between diametrically opposed bins divided by the total M$_{\star}$. Thus, the higher the asymmetry value, the more asymmetric the galaxy is. The minority of the EAGLE galaxies, 34\%, are major mergers: the ratio of the masses of the major mergers is between 1/3 and 3.

\section{Deep Learning}\label{sec:cnn}
\subsection{Convolutional Neural Networks}
Deep learning neural networks are a type of machine learning that aim to loosely mimic how a biological system processes information by using a series of layers of non-linear mathematic operations, known as neurons, each with its own weight and bias value. Here we use a type of deep learning known as convolutional neural networks (CNN). The lower layers of a CNN are known as convolutional layers and contain a defined number of kernels that are convolved with the output of the layer below. The kernels are groups of neurons that have a shape smaller in width and height than the input so must move across the input, performing the necessary matrix multiplications on the region of the input they cover at each step. The higher layers of a CNN are one-dimensional and fully connected, that is all the neurons are connected to all the neurons in the layer below. Dimensionality of the network can be reduced by applying pooling layers. This type of layer groups inputs into it and passes on the maximum or average value of each group to the next layer. When placed after an image input or a convolution layer, this grouping is done in the 2-dimensional height-width plane and not the depth/colour direction resulting in a reduction in the spatial size but not the depth. Each neuron in the kernels or fully connected layers of a network has an activation function to scale the result to pass to higher layers or force the output to a certain value, depending on the value passed into the activation function and the activation function used. The weights and biases in the neurons are trained by passing a large number of classified images through the network, in the case of supervised learning used here, such that the output classifications converge on the known input classifications. A thorough description of how CNNs work is beyond the scope of this paper; further information on CNNs can be found in \citet{lecun1998cnn}.

When discussing neural networks, some terms are used whose definitions may differ from what is expected or be unfamiliar. Also, concepts have a number of different names. To prevent confusion, terms used in this paper are defined in Table \ref{table:definitions}, taking a positive result to mean a merger and a negative result to mean a non-merger.

\begin{table*}
	\caption{Terms used when describing the performance of neural networks}
	\begin{center}
		\begin{tabular}{p{0.16\textwidth}p{0.54\textwidth}p{0.23\textwidth}}
		\hline
		Term & Definition & \\
		\hline
		Positive (P) & An object classified in the catalogues or identified by a network as a merger. & \\
		Negative (N) & An object classified in the catalogues or identified by a network as a non-merger. & \\
		True Positive (TP) & An object classified in the catalogues as a merger that is identified by a network as a merger. & \\
		False Positive (FP) & An object classified in the catalogues as a non-merger that is identified by a network as a merger. & \\
		True Negative (TN) & An object classified in the catalogues as a non-merger that is identified by a network as a non-merger. & \\
		False Negative (FN) & An object classified in the catalogues as a merger that is identified by a network as a non-merger. & \\
		Recall & Fraction of objects correctly identified by a network as a merger with respect to the total number of objects classified in the catalogues as mergers. & TP / (TP+FN) \\ 
		Fall-out & Fraction of objects incorrectly identified by a network as a merger with respect to the total number of objects classified in the catalogues as mergers. &  FP / (TP+FN) \\ 
		Specificity & Fraction of objects correctly identified by a network as a non-merger with respect to the total number of objects classified in the catalogues as non-mergers. & TN / (TN+FP) \\
		Precision & Fraction of objects correctly identified by a network as a merger with respect to the total number of objects identified by a network as a merger. & TP / (TP+FP) \\
		Negative Predictive Value (NPV) & Fraction of objects correctly identified by a network as a non-merger with respect to the total number of objects identified by a network as a non-merger. & TN / (TN+FN) \\
		Accuracy & Fraction of objects, both merger and non-merger, correctly identified by a network. & (TP+TN) / (TP+FP+TN+FN) \\
		\hline
		\end{tabular}
	\label{table:definitions}
	\end{center}
\end{table*}

\subsection{Architecture}\label{sec:architecture}
The CNN used in this work was built using the Tensorflow framework \citep{tensorflow2015-whitepaper}. As the task we are attempting to complete is similar to that of The Galaxy Challenge, we base our network on the winning \citet{2015MNRAS.450.1441D} architecture but apply some tweaks. The input image is 64 by 64 pixels with three colour channels. We then apply a series of four, two dimensional convolutional layers with 32, 64, 128 and 128 kernels of 6$\times$6, 5$\times$5, 3$\times$3 and 3$\times$3 pixels for the first, second, third and fourth layers respectively. The strides of the kernels, how far the kernel is moved as it scans the input, is set at 1 pixel for all layers and the zero padding is set to ``same'' to pad each edge of the image with zeros evenly (if required). Batch normalisation \citep{ioffe2015batch} is applied after each layer, scaling the output between zero and one, and we use Rectified Linear Units \citep[ReLU;][]{nair2010rectified} for activation. ReLU returns max($x$, 0) when passed $x$. Dropout \citep{JMLR:v15:srivastava14a} is also applied after each activation, to help reduce overfitting, with a dropout rate of 0.2, randomly setting the output of neurons to zero 20\% of the time during training. The output from the first, second and fourth convolutional layer has a 2$\times$2 pixel max-pooling applied to reduce dimensionality. After the fourth convolutional layer, we use two one-dimensional, fully connected layers of 2048 neurons, again applying ReLU activation, batch normalisation and dropout. The output layer has two neurons\footnote{It is possible to do this with a single output but this setup makes it easier to add more classes in the future.}, one for each class, and uses a softmax output. For training, validation and testing samples with an equal number of each class, as done here, softmax output provides probabilities for each class, in the interval [0, 1], that sum to one, i.e. softmax maps the un-normalised input into it into a probability distribution over the output classes. Thus there is one output that can be considered the probability the input image is of a merging system and one output that can be considered to be the probability the input image is of a non-merging galaxy. In this paper, we will use the output for the merger class, although with our binary classification the non-merger class can be considered equivalent as it is 1-(merger class output). The full network can be seen in Table \ref{table:cnn}. Loss of the network is determined using softmax cross entropy and is optimised using the Adam algorithm \citep{2014arXiv1412.6980K}. A learning rate, i.e. how fast the weights and biases in the network can change, of $5\times10^{-5}$ is used as it resulted in a more accurate network.

\begin{table*}
	\caption{Architecture of the CNN. The first column in the type of layer while the second column contains the associated properties. The input is a 64$\times$64 pixel, 3 channel image and the output is two probabilities, one for the probability the input is a merger and one for the probability the input is a non-merger. Further details on what the properties of the layers mean can be found in Sect. \ref{sec:architecture}.}
	\begin{center}
		\begin{tabular}{llllll}
		\hline
		Layer & Properties & & & & \\
		\hline
		Input & 64$\times$64 pixels & 3 channels & & & \\
		Convolutional & 32, 6$\times$6 pixel kernels & 1 pixel stride & ``same'' padding & Batch normalisation & ReLU activation\\
		Dropout & Dropout rate of 0.2 & & & & \\
		MaxPooling & 2$\times$2 pixel & 2 pixel stride & & & \\
		Convolutional & 64, 5$\times$5 pixel kernels & 1 pixel stride & ``same'' padding & Batch normalisation & ReLU activation\\
		Dropout & Dropout rate of 0.2  & & & & \\
		MaxPooling & 2$\times$2 pixel & 2 pixel stride & & & \\
		Convolutional & 128, 3$\times$3 pixel kernels & 1 pixel stride & ``same'' padding & Batch normalisation & ReLU activation\\
		Dropout & Dropout rate of 0.2  & & & & \\
		Convolutional & 128, 3$\times$3 pixel kernels & 1 pixel stride & ``same'' padding & Batch normalisation & ReLU activation\\
		Dropout & Dropout rate of 0.2  & & & & \\
		MaxPooling & 2$\times$2 pixel & 2 pixel stride & & & \\
		Flatten & & & & & \\
		Fully Connected & 2048 neurons & Batch normalisation & ReLU activation & & \\
		Dropout & Dropout rate of 0.2 & & & & \\
		Fully Connected & 2048 neurons & Batch normalisation & ReLU activation & & \\
		Dropout & Dropout rate of 0.2 & & & & \\
		Output & 2 neurons & Softmax activation & & & \\
		\hline
		\end{tabular}
	\label{table:cnn}
	\end{center}
\end{table*}

\subsection{Training, Validation and Testing}
If there are an unequal number of images in the two classes, the larger class size is reduced by randomly removing images until the classes are the same size. The images were then subdivided into three groups: 80\% were used for training, 10\% for validation and 10\% for testing. The training set was the set used to train the network while the validation was used to see how well the network was performing as training progressed. Each network was trained for 200 epochs, an epoch is showing each image to the network once, and the epoch with simultaneously the highest accuracy and lowest loss with the validation set was selected for use. Using 200 epochs is long enough as by this point the loss for the validation set has begun to increase as the network starts to over-train and learn the training set, not the features in the training set. The testing set was used once, and once only, to test the performance of the network deemed to be the best from the validation. Testing images are not used for validation to prevent accidental training on the test data set. To reduce sensitivity to galaxy orientation, the images were also augmented as they were loaded for training (and only training): the images were randomly rotated by 0$^{\circ}$, 90$^{\circ}$, 180$^{\circ}$ or 270$^{\circ}$. We also crop the images to the centre 64$\times$64 pixels and scale the images globally between zero and one, preserving the relative flux densities. The code used to create, train, validate and test the networks can be downloaded from GitHub\footnote{\label{ft:git}\url{https://github.com/wjpearson/SDSS-EAGLE-mergers}}.

\section{Results and Discussion}\label{sec:results}
We use the receiver operating characteristic (ROC) curve to determine how well the network has performed for a binary classification. The ROC curve is a plot of the recall against fall-out (see Table \ref{table:definitions} for definitions) with each point along the curve corresponding to a different value for the output (threshold) above which an input image is considered to be of a merging system. Higher recall and lower fall-out means a better threshold while the (0,0) and (1,1) positions correspond to assigning all objects to the non-merger and merger classes respectively. The threshold with recall and fall-out closest to the (1,0) position, calculated as least squared difference, is the preferred threshold for splitting mergers from non-mergers. Also, the area under the ROC curve is unity for an infallible network, and close to unity for good networks, while a truly random network will have an area of 0.5.

The two-sample Kolmogorov-Smirnov test \citep[KS-test;][]{smirnov1939estimate} is also used to compare the distributions of correctly and incorrectly identified objects to see if they are likely sampled from the same distribution. The null hypothesis that the two distributions are the same is rejected at level $\alpha$ = 0.05 if the KS-test statistic, D$_{N,M}$, is greater than Crit$_{N,M} = c(\alpha) \sqrt{\frac{n+m}{nm}}$, where $c(\alpha)$ = 1.224 for $\alpha$ = 0.05 and $n$ and $m$ are the sizes of samples $N$ and $M$.

\subsection{Observation Trained Network}\label{sec:obs_train_net}
\begin{figure}
	\centering
	\includegraphics[width=0.5\textwidth]{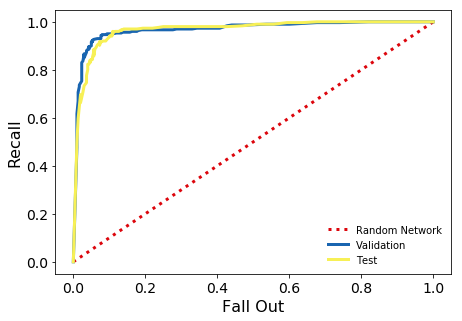}
	\caption{ROC curve for the observation network used on visually classified SDSS images at validation (blue) and testing (yellow). The area under each curve is 0.966. The dashed red line shows the position of a truly random network.}
	\label{fig:sdssroc}
\end{figure}

The 97th epoch of the network trained with SDSS images (observation network) is used. This epoch has an accuracy (see Table \ref{table:definitions} for definition) at validation of 0.932 cutting at a threshold of 0.5 to separate mergers from non-merger classification. Using the validation set, we plot the ROC curve for this network in blue in Fig. \ref{fig:sdssroc}. This has an area of 0.966 and provides an ideal cut threshold of 0.57. At this threshold, the accuracy of the validation set increases to 0.935. To determine the true accuracy of the network, we perform the same analysis for the test data set. The area under the ROC curve, shown in Fig. \ref{fig:sdssroc} in yellow, remains constant at 0.966. With the threshold set at 0.57, the final accuracy of the network is 0.915, with recall, precision, specificity and NPV of 0.920, 0.911, 0.910 and 0.919 respectively (see Table \ref{table:definitions} for definitions). It is possible to increase the accuracy, and other cut dependent statistics, by changing the cut threshold for the training set. However, this risks accidentally using the test set for training and thus not giving a true representation of the network.

Our results can be compared to those of \citet{2018MNRAS.479..415A}, who performed a similar study using the same \citet{2010MNRAS.401.1552D, 2010MNRAS.401.1043D} merger catalogue. \citet{2018MNRAS.479..415A} have a recall of 0.96, a precision of 0.97 and the area under the ROC curve is 0.9922. All these values are slightly larger than those we find, demonstrating that their network performs somewhat better. However, there are some differences between the two studies. The architecture of the CNN used here is different from that used by \citet{2018MNRAS.479..415A}, who use the Xception architecture \citep{2016arXiv161002357C}, and they perform transfer learning: using a network pre-trained on the non-astronomical ImageNet images \citep{deng2009} and then continuing to train using the merger and non-merger images. The non-merger set is also different, with \citet{2018MNRAS.479..415A} using 10\,000 non-merging galaxies as opposed to our 3003. We use an equal number of mergers and non-mergers to prevent accidental bias against the class with fewer images. Finally, the \citet{2018MNRAS.479..415A} study does not change the cut threshold from 0.5 to improve the recall or precision, suggesting that these values may be able to be improved.

Another study, by \citet{2019MNRAS.483.2968W}, trains a CNN on data from the Canada-France-Hawaii Telescope Legacy Survey \citep[CFHTLS;][]{2012AJ....143...38G}. Here, they aim to identify galaxies with tidal features, which are likely due to galaxy interactions. The performance of our SDSS trained network is much better than that of the CFHTLS network: we achieve recall of 0.920 while \citet{2019MNRAS.483.2968W} achieve 0.760. However, the differences in data set and network architecture will have an effect on the results.

To determine if certain physical properties of the galaxies are the cause of the misclassification, the specific SFR (SFR/M$_{\star}$, sSFR), M$_{\star}$, redshift and ugriz band magnitudes of the misclassified objects have been compared to their correctly classified counterparts. This will allow us to determine if, for example, all of the high mass, non-mergers have been classified as mergers. There are no trends in any of these properties: the distribution of the misclassified objects is the same as the distribution for the correctly classified objects. The confusion matrix, showing the number of TP, FP, TN and FN, for the SDSS images classified by the observation network can be found in Table \ref{tab:conf:sdss} while the KS-test statistics comparing the distributions of correctly and incorrectly identified galaxies with the physical properties can be found in Table \ref{table:ks:sdss}. See Table \ref{table:definitions} for definitions of TP, FP, TN and FN.

\begin{table}
	\caption{Confusion matrix for SDSS images classified by the observation network.}
	\begin{tabular}{c >{}r @{\hspace{0.7em}}c @{\hspace{0.4em}}c @{\hspace{0.7em}}c}
 		 & & \multicolumn{2}{c}{Network Classification} & \\
		 \multirow{3}{*}{\rotatebox[origin=c]{90}{\parbox{5cm}{\centering Catalogue Classification}}} & & Merger & Non-merger & Total \\
		& \rotatebox[origin=c]{90}{Merger} & \MyBox{276 TP}{} & \MyBox{24 FN}{} & 300 \\[2.4em]
		& \rotatebox[origin=c]{90}{Non-merger} & \MyBox{27 FP}{} & \MyBox{273 TN}{} & 300 \\[2.4em]
		& Total & 303 & 297 &
	\end{tabular}
	\label{tab:conf:sdss}
\end{table}

\begin{table}
	\caption{KS-test statistic, D$_{N,M}$, and the critical value, Crit$_{N,M} = c(\alpha) \sqrt{\frac{n+m}{nm}}$, for the SDSS images classified by the observation network. If D$_{N,M}$ > Crit$_{N,M}$, the null hypothesis that the two distributions are the same is rejected at level $\alpha$ = 0.05. Here, $c(\alpha)$ = 1.224 for $\alpha$ = 0.05 and $n$ and $m$ are the sizes of samples $N$ and $M$.}
	\begin{tabular}{ccccc}
		\hline
		Physical & & & & \\
		Parameter & D$_{TP,FN}$ & Crit$_{TP,FN}$ & D$_{TN,FP}$ & Crit$_{TN,FP}$ \\
		\hline
		M$_{\star}$ & 0.144 & 0.261 & 0.091 & 0.247 \\
		sSFR & 0.203 & 0.261 & 0.141 & 0.247 \\
		u-magnitude & 0.190 & 0.260 & 0.195 & 0.247 \\
		g-magnitude & 0.324 & 0.260 & 0.168 & 0.247 \\
		r-magnitude & 0.236 & 0.260 & 0.178 & 0.247 \\
		i-magnitude & 0.196 & 0.260 & 0.193 & 0.247 \\
		z-magnitude & 0.199 & 0.260 & 0.197 & 0.247 \\
		\hline
	\end{tabular}
	\label{table:ks:sdss}
\end{table}

\begin{figure}
	\centering
	\includegraphics[width=0.5\textwidth]{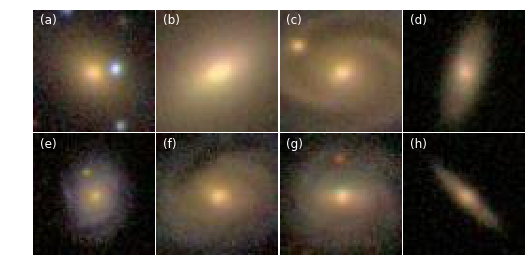}
	\caption{Examples of FP galaxies from the observation network for (a) a chance projection, (b) a galaxy filling the image, (c) a galaxy filling the image with a chance projection and (d) an isolated, non-interacting galaxy. Panels (e) to (h) show TN galaxies that are visually similar to those shown in (a) to (d).}
	\label{fig:sdss-fp}
\end{figure}

The images of the misclassified objects have also been visually inspected. Over half of the FP objects (16 of 27) have a close chance projection or a second galaxy projected into the disk of the primary galaxy, possibly fooling the network into believing that the two galaxies are merging. Four further galaxies fill the entire 64$\times$64 pixel image, two of which also have a chance projection of a second galaxy into the disk of the primary galaxy. For six of the FP, there is no clear reason why they are misclassified: they appear to be isolated galaxies without signs of morphological disturbance. The final FP is a large grand design spiral that has been identified off centre in the original 256$\times$256 pixel image. When cropped, the image contains only the arms of the spiral that appear like a disturbed system. Examples of the FP are shown in Fig. \ref{fig:sdss-fp}a-d. However, and unsurprisingly with so few misclassified objects, galaxies that are visually similar to the FP have also been correctly identified, as seen in Fig. \ref{fig:sdss-fp}e-h.

\begin{figure}
	\centering
	\includegraphics[width=0.5\textwidth]{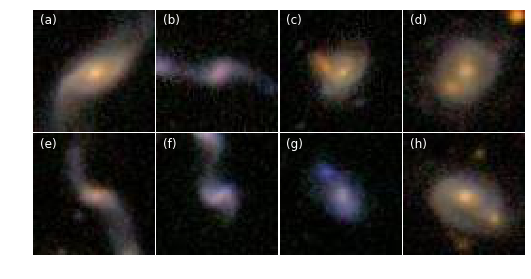}
	\caption{Examples of FN galaxies from the observation network for (a) a galaxy with its merging companion outside the image, (b) a galaxy with its merging companion on the edge of the image, (c) a merging system and (d) a merging system where the minor galaxy is almost point-like. Panels (e) to (h) show TP galaxies that are visually similar to those shown in (a) to (d).}
	\label{fig:sdss-fn}
\end{figure}

For the FN objects, six of the 24 have a merging companion that is either outside the 64$\times$64 pixel image or on the very edge, indicating that a larger image may reduce the FP rate. The remaining images show a clear morphological disturbance or a clear merger companion. It is possible that these companions are being identified by the network as chance projections, especially the companions that are almost point-like in the image. Examples of the FN are shown in Fig. \ref{fig:sdss-fn}a-d. As with the FP objects, there are also example TP that are visually similar to the FN galaxies, presented in Fig. \ref{fig:sdss-fn}e-h.

\subsection{Simulation Trained Network}\label{sec:sim_train_net}
\begin{figure}
	\centering
	\includegraphics[width=0.5\textwidth]{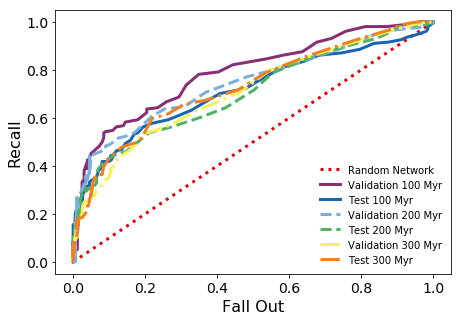}
	\caption{ROC curve for the simulation networks at validation (purple, light blue, yellow) and testing (bark blue, green, orange) for 100~Myr (solid), 200~Myr (dashed) and 300~Myr (dot-dashed) from the merger event. The areas under the curves can be found in Table \ref{table:stats}. The dashed red line shows the position of a truly random network.}
	\label{fig:eagleroc}
\end{figure}

The 26th epoch of the network trained with EAGLE images (simulation network) is used. This epoch has an accuracy of 0.672 at validation, cutting at a threshold of 0.5 to separate mergers from non-merger classification. Using the validation set, we plot the ROC curve for this network in dot-dashed yellow in Fig. \ref{fig:eagleroc}. This has an area of 0.710 and provides an ideal cut threshold of 0.46. At this threshold, the accuracy of the validation set decreases to 0.644. To determine the true accuracy of the network, we perform the same analysis for the test data set. The area under the ROC curve, the dot-dashed orange curve in Fig. \ref{fig:eagleroc}, increases to 0.726. With the threshold set at 0.46, the final accuracy of the network is 0.674, with recall, precision, specificity and NPV of 0.657, 0.680, 0.692 and 0.668 respectively.

The lower accuracy of the simulation trained network relative to the observation trained network (discussed in Sect. \ref{sec:obs_train_net}) is a result of the difference in the training sample. The SDSS merger sample has been thoroughly checked to verify there are visible indications of a merger, as can been seen in the examples in Fig. \ref{fig:sdssimg}. The EAGLE merger sample, however, contains physically classified mergers in the simulation without visual inspection to check whether there are any obvious signs of merging. As such, the EAGLE merger sample includes a wide variety of merger types (in terms of their mass ratios, orbital parameters, gas fractions, etc.) and hence some of the mergers are bound to have inconspicuous merging signs and will therefore be harder to discern, resulting in a lower accuracy. We have checked the merging galaxies misclassified by the network visually and confirm that most EAGLE mergers are indeed not as conspicuous as the ones in the SDSS catalogue, where the mergers from \citet{2010MNRAS.401.1552D, 2010MNRAS.401.1043D} have been selected to be conspicuous.

If the time before and after the merger event is decreased, the accuracy of the network increases. Performing the same analysis as with the full EAGLE data set, we find that using galaxies that are within 200~Myr of the merger event results in a network that has a test accuracy of 0.644 at a cut threshold of 0.40 in the 31st epoch. Using galaxies that are within 100~Myr of the merger event the network as a test accuracy of 0.652 at a cut threshold of 0.39 in the 52nd epoch. The full statistics for these two networks can be found in Table \ref{table:stats} and the ROC curves can be found in Fig. \ref{fig:eagleroc}. As the 100~Myr network has the largest area under the ROC curve, the majority of the remainder of the paper will now focus on the 100~Myr network when discussing the simulation network. The confusion matrix for the 100~Myr network can be found in Table \ref{tab:conf:eagle}.

\begin{table}
	\caption{Statistics for the SDSS and the 100~Myr, 200~Myr and 300~Myr EAGLE trained networks at testing.}
	\begin{center}
		\begin{tabular}{lllll}
		\hline
		 & SDSS & 100~Myr & 200~Myr & 300~Myr\\
		\hline
		Epoch used   &      97 &      52 &      31 &     26\\
		Cut threshold & 0.57  &    0.39 &   0.40 &  0.46\\
		ROC area      & 0.966 & 0.787 & 0.704 & 0.726\\
		Recall            & 0.920 & 0.632 & 0.562 & 0.657\\
		Precision       & 0.911 & 0.658 & 0.673 & 0.680\\
		Specificity      & 0.910 & 0.672 & 0.726 & 0.692\\
		NPV               & 0.919 & 0.646 & 0.624 & 0.668\\
		Accuracy       & 0.915 & 0.652 & 0.644 & 0.674\\
		\hline
		\end{tabular}
	\label{table:stats}
	\end{center}
\end{table}

\begin{table}
	\caption{Confusion matrix for EAGLE images classified by the simulation network.}
	\begin{tabular}{c >{}r @{\hspace{0.7em}}c @{\hspace{0.4em}}c @{\hspace{0.7em}}c}
 		 & & \multicolumn{2}{c}{Network Classification} & \\
		 \multirow{3}{*}{\rotatebox[origin=c]{90}{\parbox{5cm}{\centering Catalogue Classification}}} & & Merger & Non-merger & Total \\
		& \rotatebox[origin=c]{90}{Merger} & \MyBox{127 TP}{} & \MyBox{74 FN}{} & 201 \\[2.4em]
		& \rotatebox[origin=c]{90}{Non-merger} & \MyBox{66 FP}{} & \MyBox{135 TN}{} & 201 \\[2.4em]
		& Total & 193 & 209 &
	\end{tabular}
	\label{tab:conf:eagle}
\end{table}

A similar study has been performed by \citet{2018arXiv180902136S} using simulated galaxy images from the Illustris simulation \citep{2014MNRAS.444.1518V}, although their technique is somewhat different. In their study, \citet{2018arXiv180902136S} train Random Forests using non-parametric morphology statistics, such as concentration, asymmetry, Gini and M$_{20}$, as inputs, with these statistics derived from Illustris galaxies processed to look like Hubble Space Telescope images. They select galaxies that will, or have, merge within 250~Gyr. The recall of \citet{2018arXiv180902136S} is slightly higher than this work, they achieve $\approx 0.70$ compared to our 0.632, but their precision is much lower, at $\approx 0.30$ compared to 0.658. Comparing the \citet{2018arXiv180902136S} results to our 300~Myr trained network, a more fair comparison, shows similar results: \citet{2018arXiv180902136S} has higher recall, $\approx 0.70$ compared to 0.657, but lower precision, $\approx 0.30$ compared to 0.680.

As the galaxies are generated from a simulation, we know the physical properties of these systems. As with the SDSS objects, we can compare the physical properties of the galaxies that are correctly and incorrectly identified. KS-test statistics comparing the distributions of correctly and incorrectly identified galaxies with the physical properties can be found in Table \ref{table:ks:eagle}.

\begin{figure}
	\centering
	\includegraphics[width=0.5\textwidth]{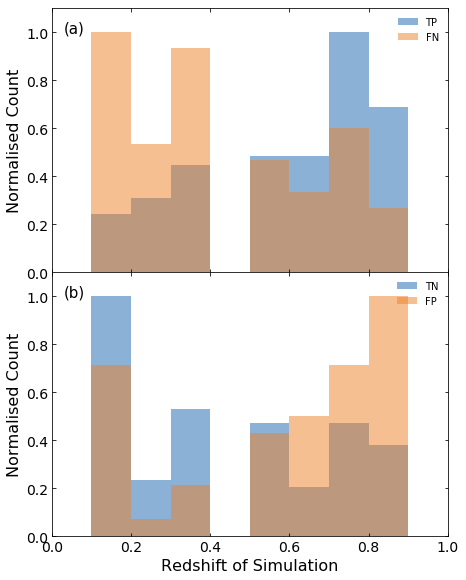}
	\caption{Distributions for the correctly (blue) and incorrectly (orange) identified EAGLE objects in the simulation network for (a) mergers and (b) non-mergers as a function of simulation snapshot redshift. Merging objects with low snapshot redshifts are disproportionally assigned a non-merger classification while non-merging objects with high simulation redshifts are often seen as mergers.}
	\label{fig:eaglez}
\end{figure}

Many FN objects appear to have low simulation snapshot redshifts when compared to the TP, see Fig. \ref{fig:eaglez}a, potentially a result of coarser time resolution of the simulation at low redshift. Note that the simulation snapshot redshift is different from the redshift used when making the EAGLE galaxies look like SDSS images. For the TN and FP populations, higher snapshot redshifts have a higher fraction of FP sources relative to the TN, see Fig. \ref{fig:eaglez}b. This suggests that simulated non-mergers in the local universe look different from simulated non-mergers in the higher-z universe.

The asymmetry of the non-merger population also has an effect: non-merging objects with higher asymmetry are preferentially being identified as merging systems. It is worth noting that the time to/from the merger event does not appear directly correlated with the asymmetry of the galaxy. 

\begin{table}
	\caption{KS-test statistic, D$_{N,M}$, and the critical value, Crit$_{N,M} = c(\alpha) \sqrt{\frac{n+m}{nm}}$, for the EAGLE images classified by the simulation network. If D$_{N,M}$ > Crit$_{N,M}$, the null hypothesis that the two distributions are the same is rejected at level $\alpha$ = 0.05. Here, $c(\alpha)$ = 1.224 for $\alpha$ = 0.05 and $n$ and $m$ are the sizes of samples $N$ and $M$.}
	\begin{tabular}{ccccc}
		\hline
		Physical & & & & \\
		Parameter & D$_{TP,FN}$ & Crit$_{TP,FN}$ & D$_{TN,FP}$ & Crit$_{TN,FP}$ \\
		\hline
		Projection & \multirow{2}{*}{0.102} & \multirow{2}{*}{0.179} & \multirow{2}{*}{0.162} & \multirow{2}{*}{0.184} \\
		Redshift & & & & \\
		Simulation & \multirow{2}{*}{0.272} & \multirow{2}{*}{0.179} & \multirow{2}{*}{0.260} & \multirow{2}{*}{0.184} \\
		Redshift & & & & \\
		Asymmetry & 0.225 & 0.179 & 0.232 & 0.184 \\
		Time since & \multirow{2}{*}{0.123} & \multirow{2}{*}{0.179} & \multirow{2}{*}{-} & \multirow{2}{*}{-} \\
		Merger & & & & \\
		u-magnitude & 0.286 & 0.179 & 0.120 & 0.184 \\
		g-magnitude & 0.259 & 0.179 & 0.075 & 0.184 \\
		r-magnitude & 0.222 & 0.179 & 0.068 & 0.184 \\
		i-magnitude & 0.216 & 0.179 & 0.065 & 0.184 \\
		z-magnitude & 0.201 & 0.179 & 0.072 & 0.184 \\
		Mass ratio & 0.164 & 0.179 & - & - \\
		M$_{\star}$ & 0.225 & 0.179 & 0.205 & 0.184 \\
		sSFR & 0.338 & 0.179 & 0.207 & 0.191 \\
		\hline
	\end{tabular}
	\label{table:ks:eagle}
\end{table}

\begin{figure}
	\centering
	\includegraphics[width=0.5\textwidth]{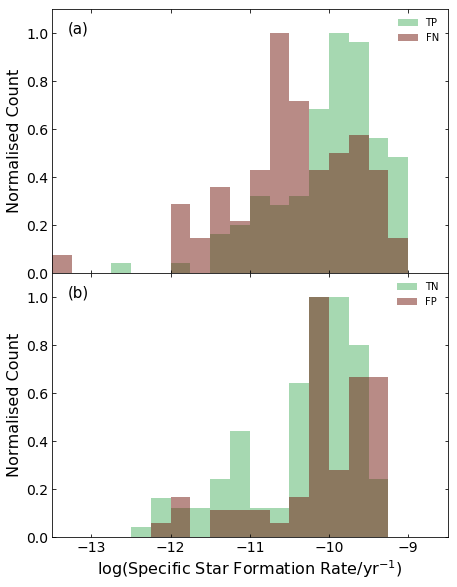}
	\caption{Distribution of EAGLE galaxies from the simulation network of the correctly (green) and incorrectly (brown) mergers (a) and non-mergers (b) as a function of EAGLE sSFR. Merging galaxies with low sSFR are often misclassified as non-merging while high sSFR non-mergers are often identified as mergers.}
	\label{fig:eagle-sfr}
\end{figure}

In M$_{\star}$, there is a slight trend for the low mass, merging systems to be identified as non-mergers, although the non-merging galaxies are typically slightly lower mass than the merging systems so this is not overly unexpected. For sSFR, there is a splitting, with low sSFR merging systems being preferentially assigned the non-merger classification and the high sSFR non-merging galaxies preferentially identified as mergers, as shown in Fig. \ref{fig:eagle-sfr}.

\begin{figure}
	\centering
	\includegraphics[width=0.5\textwidth]{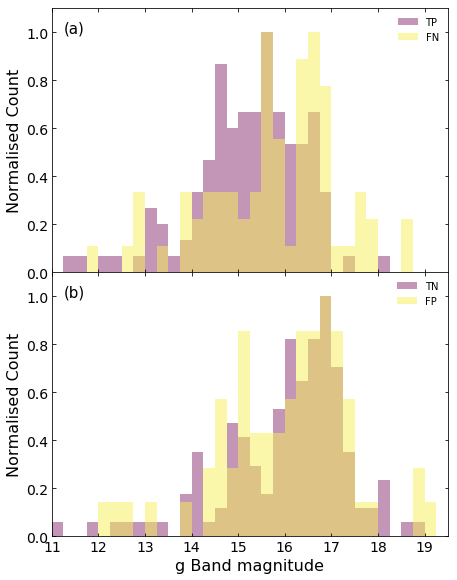}
	\caption{Distributions for the correctly (purple) and incorrectly (yellow) identified objects for (a) mergers and (b) non-mergers as a function of g-band magnitude for EAGLE galaxies classified by the simulation network. Faint mergers are preferentially classified as non-mergers while the distribution of misclassified non-mergers is at intermediate magnitudes.}
	\label{fig:eagle-mag}
\end{figure}

The apparent magnitude of the simulated galaxy after redshift projection has the largest effect, compared to the other parameters investigated, on the correct identification. For the merging systems, faint objects are preferentially classified as non-merging systems while the bright non-mergers are more likely to be misclassified as mergers. An example of this in the g-band is presented in Fig. \ref{fig:eagle-mag}. Misclassification for the merging systems is likely a result of the merging systems being brighter, on average, than the non-merger systems while the majority of the merging systems are fainter, hence the high misclassification rate for non-merging systems at these magnitudes.

\begin{figure}
	\centering
	\includegraphics[width=0.5\textwidth]{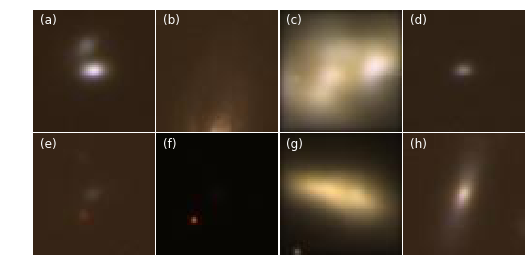}
	\caption{Examples of FP EAGLE galaxies from the simulation network for (a) a chance projection, (b) a galaxy where the chance projection from the SDSS noise has resulted in the EAGLE galaxy appearing faint in the image, (c) a galaxy at low projection redshift and (d) an isolated, non-interacting galaxy. Panels (e) to (h) show TN galaxies that are visually similar to those shown in (a) to (d).}
	\label{fig:eagle-fp}
\end{figure}

As with the SDSS images, the misclassified EAGLE images have also been visually inspected. The majority of the FP galaxies, 43 of 66, contain a chance projection generated when the real SDSS noise is added. Three FP galaxies have a projected galaxy that is much brighter than the EAGLE galaxy, resulting in the EAGLE galaxy becoming extremely faint in the image and (almost) impossible to see by eye. There are also correctly identified galaxies that also suffer from the same image suppression, suggesting that this issue is not the sole cause of the misclassification. Of the remaining FP, three are at a low projection redshift, resulting in the features and inhomogeneities of the galaxy appearing as morphological disturbances, although, again, there are examples of these low projection redshift galaxies that have been correctly identified as non-merging. The other objects show no signs of asymmetry or morphological disturbances. Examples of these galaxies can be found in Fig. \ref{fig:eagle-fp}a-d while example TN galaxies that are visually similar can be found in Fig. \ref{fig:eagle-fp}e-h.

\begin{figure}
	\centering
	\includegraphics[width=0.5\textwidth]{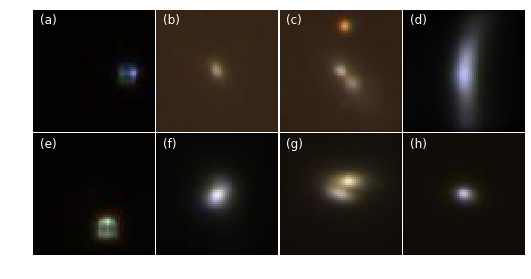}
	\caption{Examples of FN EAGLE galaxies from the simulation network for (a) a galaxy where the chance projection from the SDSS noise has resulted in the EAGLE galaxy appearing faint in the image, (b) a merging system that appears as a single, smooth galaxy, (c) a galaxy with a clearly identifiable counterpart and (d) as asymmetric galaxy. Panels (e) to (h) show TP galaxies that are visually similar to those shown in (a) to (d).}
	\label{fig:eagle-fn}
\end{figure}

For the FN objects, 9 of the 74 have a bright chance projection from the added SDSS noise that results in the EAGLE galaxy becoming (almost) impossible to see in the image. As these types of images are present in the FN, FP, TP and TN, it is unlikely that the bright counterpart is causing the misclassifications. 21 FNs do not appear morphologically disturbed or asymmetric. This is likely a result of the PSF convolution and redshift re-projection smoothing out the visual merger indicators resulting in what appears to be a single, smooth galaxy. The remaining objects do have clearly identifiable merger counterparts or asymmetry. Examples of these galaxies can be found in Fig. \ref{fig:eagle-fn}a-d. As with the EAGLE FP, Fig. \ref{fig:eagle-fn}e-h show there are also examples of visually similar galaxies that have been correctly identified by the simulation network.

The CNN architecture was also trained on simulation images that had only been partially processed to look like SDSS images. This will allow us to determine if there is a specific part of the process that results in the lower accuracy for the simulation network with respect to the observation network. For this, we use EAGLE galaxies that are within 100~Myr of the merger event and perform one of the following processes: convolve the EAGLE image with the SDSS PSF (C), inject the EAGLE image into the real SDSS noise (N), match the EAGLE resolution to that of the SDSS images (R), adjust the EAGLE magnitude to be the correct apparent magnitude for a chosen redshift (Z) or a combination of three (CNR, CNZ, CRZ, NRZ). We also train the network on the EAGLE images that have not been processed. As with training with SDSS or fully processed EAGLE images, the epoch with simultaneously the lowest validation loss and accuracy is chosen and the cut threshold with fall-out and recall closet to (0,1) is used. The statistics are then calculated for the test set and are presented in Table \ref{table:mangle}. Individually, C and N do not notably change the accuracy of the trained network, remaining within one percentage point of the accuracy of the un-processed EAGLE images (87\%). R and Z both cause a small reduction in the accuracy, reducing it to 82\%, along with the combination of CNR. CNZ, CRZ and NRZ all show more significant reductions. Of these three, CRZ has the smallest effect, reducing accuracy to 78\%, while the combinations CNZ and NRZ reduce the accuracy to 72\% and 68\% respectively. This suggests that correcting the apparent magnitude has the largest affect, especially when coupled with the inclusion of back ground noise. This is possibly because when changing from absolute to apparent magnitude, the fainter objects are becoming harder to discern from the background when injected into the real SDSS noise. We note, however, that only 58 of the original 10\,134 processed EAGLE images have an apparent r-band magnitude greater than the limit applied to the SDSS images.

\begin{table*}
	\caption{Statistics for the network trained with partially processed EAGLE images at testing. C is convolving the EAGLE image with the SDSS PSF, N is injecting the EAGLE image into the real SDSS noise, R is matching the EAGLE resolution to that of SDSS and Z is changing the EAGLE magnitude to apparent from absolute.}
	\begin{center}
		\begin{tabular}{llllllllll}
		\hline
		Processing    & None &       C &       N &       R &       Z &    CNR &   CNZ &   CRZ & NRZ \\
		\hline
		Epoch used   &   183 &     167 &    191 &    135 &   116 &     167 &      43 &    145 &     36 \\
		Cut threshold &  0.47 &    0.48 &   0.42 &   0.45 &  0.47 &    0.48 &   0.40 &   0.50 &  0.43 \\
		ROC area      & 0.943 & 0.948 & 0.933 & 0.903 & 0.893 & 0.911 & 0.803 & 0.866 & 0.763 \\
		Recall            & 0.841 & 0.841 & 0.866 & 0.786 & 0.791 & 0.776 & 0.692 & 0.667 & 0.657 \\
		Precision       & 0.904 & 0.889 & 0.888 & 0.845 & 0.846 & 0.867 & 0.735 & 0.859 & 0.688 \\
		Specificity      & 0.910 & 0.896 & 0.891 & 0.856 & 0.856 & 0.881 & 0.751 & 0.889 & 0.701 \\
		NPV               & 0.851 & 0.849 & 0.869 & 0.800 & 0.804 & 0.797 & 0.709 & 0.725 & 0.671 \\
		Accuracy       & 0.876 & 0.868 & 0.866 & 0.821 & 0.823 & 0.828 & 0.721 & 0.778 & 0.679 \\
		\hline
		\end{tabular}
	\label{table:mangle}
	\end{center}
\end{table*}

\subsection{Cross application of the networks}
Here we pass the images through the other network, that is we pass all 6006 SDSS images through the simulation network and all 4020 EAGLE images through the observation network. For this, we use the same cut threshold as for passing through the ``correct'' images. This is done so that we can understand any biases and incompleteness in the two data sets. For example, the visually classified mergers from the SDSS data consist only of certain types of mergers with conspicuous merging signs, e.g., two massive galaxies obviously interacting with strong tidal features. However, the EAGLE simulation contains a much more complete merger sample. So, one would expect the neural network trained with the visually classified SDSS merger sample to perform poorly on simulated images of EAGLE mergers. We also perform the cross application so that any SDSS objects classified as merging systems in the visual classification but not identified by the simulation trained network can be identified and help improve our understanding of the limitations of simulations so that they become more representative of the real universe in future developments.

\subsubsection{EAGLE images through the observation network}\label{sec:eagle_t_sdss}
\begin{figure}
	\centering
	\includegraphics[width=0.5\textwidth]{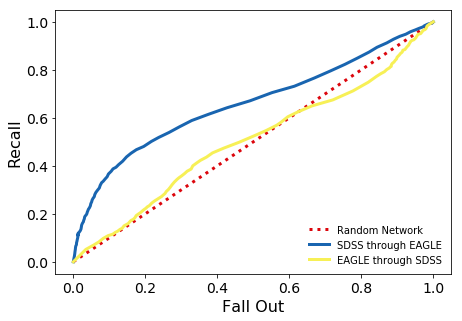}
	\caption{ROC curve for the SDSS images classified by the simulation network (blue) and the EAGLE images classified by the observation network (yellow). The area under the EAGLE through observation network is 0.502 while the area under the SDSS through simulation network is 0.689. The dashed red line shows the position of a truly random network.}
	\label{fig:wrongROC}
\end{figure}

Passing all the EAGLE images through the observation network resulted in an accuracy of 0.530, only slightly better than random assignment of objects at first glance. Precision and NPV are similarly close to random at 0.541 and 0.523. However, recall is low, at 0.387, and the specificity is high, at 0.673, demonstrating that the network preferentially assigns objects to the non-merger class but with each class containing just over half correctly identified objects. As to be expected, the area under the ROC curve is close to 0.5 at 0.502, depicted in yellow in Fig. \ref{fig:wrongROC}. The confusion matrix can be found in Table \ref{tab:conf:eagle-t-sdss}.

\begin{table}
	\caption{Confusion matrix for EAGLE images classified by the observation network.}
	\begin{tabular}{c >{}r @{\hspace{0.7em}}c @{\hspace{0.4em}}c @{\hspace{0.7em}}c}
 		 & & \multicolumn{2}{c}{Network Classification} & \\
		 \multirow{3}{*}{\rotatebox[origin=c]{90}{\parbox{5cm}{\centering Catalogue Classification}}} & & Merger & Non-merger & Total \\
		& \rotatebox[origin=c]{90}{Merger} & \MyBox{777 TP}{} & \MyBox{1352 FN}{} & 2010 \\[2.4em]
		& \rotatebox[origin=c]{90}{Non-merger} & \MyBox{658 FP}{} & \MyBox{1233 TN}{} & 2010 \\[2.4em]
		& Total & 1435 & 2585 &
	\end{tabular}
	\label{tab:conf:eagle-t-sdss}
\end{table}

\begin{figure}
	\centering
	\includegraphics[width=0.5\textwidth]{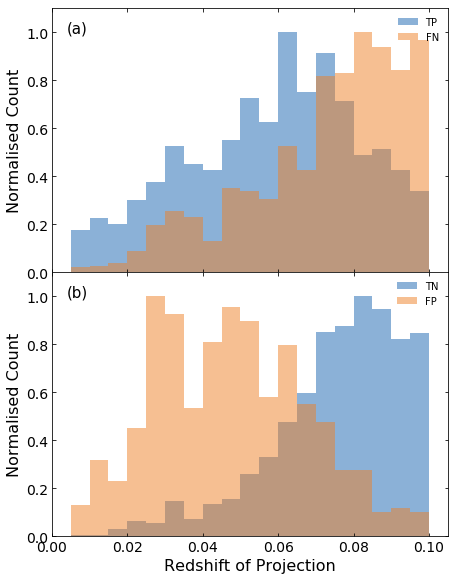}
	\caption{Distributions for the correctly (blue) and incorrectly (orange) identified EAGLE objects for (a) mergers and (b) non-mergers as a function of redshift used for projection after being classified by the observation network. High redshift, merging systems are preferentially classified as non-merging while low redshift non-merging systems are preferentially classified as merging.}
	\label{fig:eagle-t-sdss-z}
\end{figure}

As before, the physical properties of the EAGLE images can be examined to determine if they are affecting the classification by the network, a brief summary of which can be seen in the KS-test results in Table \ref{table:ks:eagle-t-sdss}. One property that has an obvious splitting between correct and incorrect assignment is the redshift of projection. As is evident in Fig. \ref{fig:eagle-t-sdss-z}, objects with high projection redshifts are preferentially being classified as non-merger systems, see Fig. \ref{fig:eagle-t-sdss-z}a, while objects with low projected redshifts are classified as merging systems, see Fig. \ref{fig:eagle-t-sdss-z}b. The distribution of redshifts used to re-project the EAGLE galaxies is nearly identical to the SDSS distribution: the redshifts used to re-project the galaxies were drawn randomly from the redshifts of the SDSS observations. Thus this effect is not a result of a mismatch in the redshift distributions between observations and simulations. The issue of misclassified mergers at high redshift may arise while matching the physical resolution (i.e. kpc per pixel) of the EAGLE images to the SDSS images. At high redshift, this could result in a loss of finer detail that would be expected in merging systems, resulting in these systems being classified as non-mergers. The main misclassification of non-merging systems happens at low projection redshift. The physical resolution of EAGLE images matches the physical resolution of the SDSS images at $z \approx 0.03$. Objects assigned a redshift lower than this value are increased in physical resolution using a bicubic interpolation. This interpolation may result in the creation of artifacts that appear, to the CNN, like features of merging systems. Alternatively, it is possible that at low redshifts the individual particles of the simulation are detectably disturbing the light profile of the galaxies and resulting in misclassification.

\begin{table}
	\caption{KS-test statistic, D$_{N,M}$, and the critical value, Crit$_{N,M} = c(\alpha) \sqrt{\frac{n+m}{nm}}$, for the EAGLE images classified by the observation network. If D$_{N,M}$ > Crit$_{N,M}$, the null hypothesis that the two distributions are the same is rejected at level $\alpha$ = 0.05. Here, $c(\alpha)$ = 1.224 for $\alpha$ = 0.05 and $n$ and $m$ are the sizes of samples $N$ and $M$.}
	\begin{tabular}{ccccc}
		\hline
		Physical & & & & \\
		Parameter & D$_{TP,FN}$ & Crit$_{TP,FN}$ & D$_{TN,FP}$ & Crit$_{TN,FP}$ \\
		\hline
		Projection & \multirow{2}{*}{0.306} & \multirow{2}{*}{0.056} & \multirow{2}{*}{0.5812} & \multirow{2}{*}{0.058} \\
		Redshift & & & & \\
		Simulation & \multirow{2}{*}{0.100} & \multirow{2}{*}{0.056} & \multirow{2}{*}{0.086} & \multirow{2}{*}{0.058} \\
		Redshift & & & & \\
		Asymmetry & 0.116 & 0.056 & 0.050 & 0.058 \\
		Time since & \multirow{2}{*}{0.152} & \multirow{2}{*}{0.056} & \multirow{2}{*}{-} & \multirow{2}{*}{-} \\
		Merger & & & & \\
		u-magnitude & 0.412 & 0.056 & 0.562 & 0.058 \\
		g-magnitude & 0.436 & 0.056 & 0.605 & 0.058 \\
		r-magnitude & 0.416 & 0.056 & 0.612 & 0.058 \\
		i-magnitude & 0.403 & 0.056 & 0.612 & 0.058 \\
		z-magnitude & 0.394 & 0.056 & 0.607 & 0.058 \\
		Mass ratio & 0.074 & 0.056 & - & - \\
		M$_{\star}$ & 0.199 & 0.056 & 0.127 & 0.058 \\
		sSFR & 0.110 & 0.056 & 0.049 & 0.060 \\
		\hline
	\end{tabular}
	\label{table:ks:eagle-t-sdss}
\end{table}

There is also a trend with the mass ratio of the merging systems. Although the TP and FN do not split into two distinct distributions, the low mass ratio merger systems, i.e. major mergers, are more often misclassified as non-merging galaxies. This is the opposite to what would be expected: minor mergers would be expected to be misclassified more often as the disturbances from the smaller galaxy would be expected to be less obvious. Similarly, low mass mergers have a slight preference to be assigned the non-merger class and vice versa, although again this is unsurprising as the merger sample is typically higher mass than the non-merger sample for the EAGLE galaxies. The mass ranges for the EAGLE and SDSS data are not quite comparable: while the SDSS data has an effective mass limit of $10^{10}$~M$_{\odot}$ at $z=0.1$, there are lower mass galaxies in the sample, while the mass limit for EAGLE is $10^{10}$~M$_{\odot}$ at all projected redshifts. For the sSFR, the high sSFR merging galaxies are often misclassified as non-mergers while there is no obvious misclassification for the non-merging galaxies.

\begin{figure}
	\centering
	\includegraphics[width=0.5\textwidth]{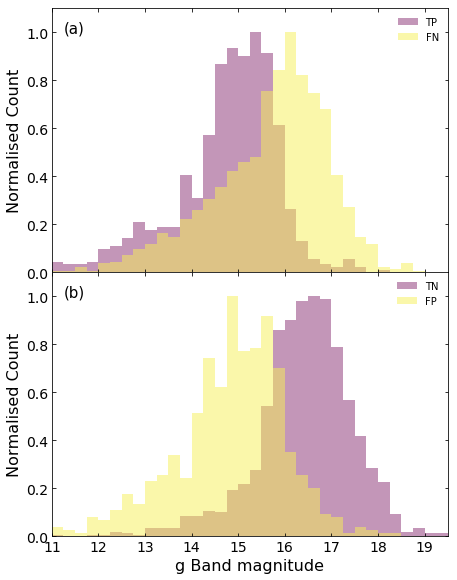}
	\caption{Distributions for the correctly (purple) and incorrectly (yellow) identified EAGLE objects for (a) mergers and (b) non-mergers as a function of g-band magnitude after being classified by the observation network. Faint, merging systems are preferentially classified as non-merging while bright non-merging systems are preferentially classified as merging.}
	\label{fig:eagle-t-sdss-mag}
\end{figure}

As with the EAGLE images through the simulation network, the apparent magnitude of the object has the largest effect on the correct classification. Like the simulation network, the observation network also identifies faint, merging systems as non-merging systems and identifies bright, non-merging systems as merging systems. This is true for all five of the ugriz bands. An example in the g-band is presented in Fig. \ref{fig:eagle-t-sdss-mag}. This is consistent with the results seen with the projected redshift above and is likely a result of the re-projection to the projection redshift.

It is also more likely that the more complete classification for the EAGLE galaxies is causing the low accuracy. The SDSS classifications are for objects that are clearly visually merging systems while the EAGLE classifications will include systems that are not obviously, visually merging. Thus, the observation trained network has not been trained to identify merging systems that are not obviously, visually merging and hence assign these objects the non-merger classification, increasing the number of FN.

\begin{figure}
	\centering
	\includegraphics[width=0.5\textwidth]{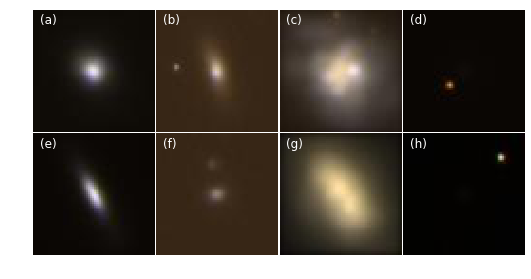}
	\caption{Examples of EAGLE FP galaxies (a to d) from the observation network for (a) an isolated, non-interacting galaxy, (b) a chance projection, (c) a galaxy at low projection redshift and (d) a galaxy where the chance projection from the SDSS noise has resulted in the EAGLE galaxy appearing faint in the image. Panels (e) to (h) show TN galaxies that are visually similar to those shown in (a) to (d).}
	\label{fig:eagle_t_sdss-fp}
\end{figure}

Visual inspection of a subsample of the FP shows that the majority of these objects ($\approx$64\%) appear to be isolated, non-interacting systems. A further $\approx$18\% of the objects have a close chance projection that may be being mistaken for a merging partner by the CNN. $\approx$8\% of the FP objects are galaxies that have been projected into a larger angular size than the original, raw image from EAGLE. This often results in the internal structure of the galaxy being expanded and could appear to the network to be morphological disturbances or multiple galaxies. The remaining objects have a bright chance projection in the SDSS noise and, as a result, are (almost) impossible to see in the image. Examples of these galaxies can be found in Fig. \ref{fig:eagle_t_sdss-fp}a-d. With accuracy, specificity and NPV all being almost equivalent to 0.5, it is unsurprising to find examples of visually similar galaxies that have been correctly identified and are shown in Fig. \ref{fig:eagle_t_sdss-fp}e-h.

\begin{figure}
	\centering
	\includegraphics[width=0.5\textwidth]{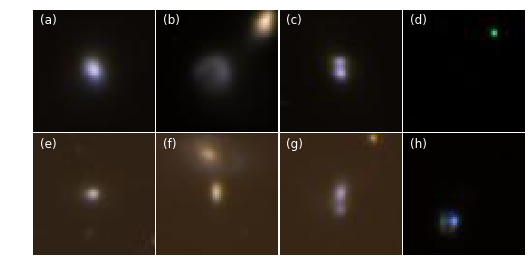}
	\caption{Examples of EAGLE FN galaxies (a to d) from the observation network for (a) an apparent single object, (b) a galaxy with a counterpart, either a merger counterpart from EAGLE or a chance projection from the SDSS noise, (c) an unambiguous merger and (d) a galaxy where the chance projection from the SDSS noise has resulted in the EAGLE galaxy appearing faint in the image. Panels (e) to (h) show TP galaxies that are visually similar to those shown in (a) to (d).}
	\label{fig:eagle_t_sdss-fn}
\end{figure}

Images of the FN are more useful in understanding why the EAGLE images are poorly classified by the observation network. Inspecting a subsample, nearly half ($\approx$46\%) appear to be a single object. This suggests that these objects have had the visible signatures of merger suppressed while being processed to look like SDSS images, likely by the re-projection and PSF matching, or that these mergers are not obvious, even without the processing steps to make them look like SDSS images. It could also be that the merging companion is hidden behind the galaxy it is merging with, as the angle the galaxy is viewed at is picked randomly, so it cannot be seen within the image. Of the remaining objects, $\approx$41\% had at least one counterpart, either from the simulation or random projections from the SDSS noise, that could potentially be merging with the central galaxy and $\approx$5\% were unambiguously merging systems. As with the FP, there are a number of images whose simulated galaxies have been suppressed by bright chance projections from the SDSS noise. Example FN galaxies can be found in Fig. \ref{fig:eagle_t_sdss-fn}a-d and their visually similar but correctly identified counterparts can be found in Fig. \ref{fig:eagle_t_sdss-fn}e-h.

There are limitations of the cross application due to the selection of the EAGLE galaxies. To increase the number of galaxies available to train the simulation network we use merging and non-merging galaxies with simulation redshifts out to $z=1$ and re-project them to redshifts between 0.005 and 0.1. However, the EAGLE galaxies at $z=1$ have approximately half the radii of a galaxy at $z=0$ while the radii of EAGLE galaxies is similar to those observed in the real universe \citep{2017MNRAS.465..722F}. As a result, images of the EAGLE galaxies can contain an object up to 2 times too small to be comparable in size to the observations, which may hamper the observation network's ability to correctly identify merging and non-merging systems. If this limitation was reduced, or removed entirely, it is possible that the results of this cross application could improve.

\subsubsection{SDSS images through the simulation network}
Passing all the SDSS images through the simulation network was more successful than passing all the EAGLE images through the observation network. While still not as good as SDSS images through the observation network, the SDSS images classified by the simulation network had an accuracy of 0.646. Like the EAGLE images through the observation network, the SDSS images through the simulation network have a preference towards the non-merger assignment, demonstrated by a low recall of 0.467 and high specificity of 0.825. The area under the ROC curve is 0.658, see the blue line in Fig. \ref{fig:wrongROC}. The statistics for the cross application of the networks can be found in Table \ref{table:wrong}. The confusion matrix, showing the number of correctly and incorrectly identified objects, can be found in Table \ref{tab:conf:sdss-t-eagle}.

\begin{table}
	\caption{Statistics for the EAGLE images classified by the observation network and the SDSS images classified by the simulation network.}
	\begin{center}
		\begin{tabular}{lll}
		\hline
		Images & EAGLE & SDSS \\
		\hline
		Network & Observation & Simulation \\
		\hline
		Cut threshold & 0.57  &    0.39\\
		ROC area      & 0.502 & 0.658\\
		Recall            & 0.387 & 0.467\\
		Precision       & 0.541 & 0.727\\
		Specificity      & 0.673 & 0.825\\
		NPV               & 0.523 & 0.608\\
		Accuracy       & 0.530 & 0.646\\
		\hline
		\end{tabular}
	\label{table:wrong}
	\end{center}
\end{table}

\begin{table}
	\caption{Confusion matrix for SDSS images classified by the simulation network.}
	\begin{tabular}{c >{}r @{\hspace{0.7em}}c @{\hspace{0.4em}}c @{\hspace{0.7em}}c}
 		 & & \multicolumn{2}{c}{Network Classification} & \\
		 \multirow{3}{*}{\rotatebox[origin=c]{90}{\parbox{5cm}{\centering Catalogue Classification}}} & & Merger & Non-merger & Total \\
		& \rotatebox[origin=c]{90}{Merger} & \MyBox{1403 TP}{} & \MyBox{1600 FN}{} & 3003 \\[2.4em]
		& \rotatebox[origin=c]{90}{Non-merger} & \MyBox{526 FP}{} & \MyBox{2477 TN}{} & 3003 \\[2.4em]
		& Total & 1929 & 4077 &
	\end{tabular}
	\label{tab:conf:sdss-t-eagle}
\end{table}

\begin{figure}
	\centering
	\includegraphics[width=0.5\textwidth]{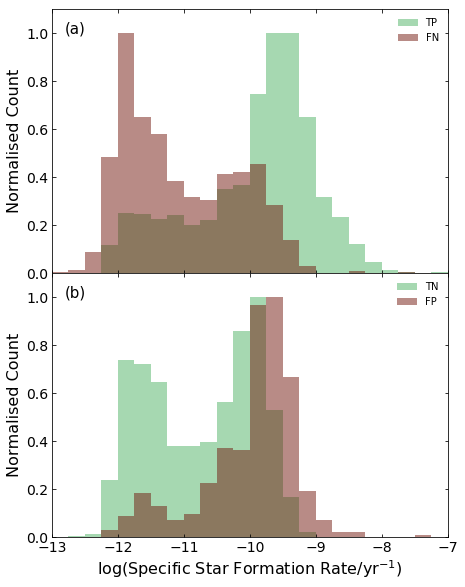}
	\caption{Distributions for the correctly (green) and incorrectly (brown) identified SDSS objects for (a) mergers and (b) non-mergers as a function of sSFR after being classified by the simulation network. Low sSFR merging systems are preferentially classified as non-merging while high sSFR non-merging systems have a higher misclassification rate than low sSFR non-merger.}
	\label{fig:sdss-t-eagle-sfr}
\end{figure}

As with the SDSS images identified by the observation network, we can examine the estimated physical parameters of the SDSS images that were classified by the simulation network. The KS-test statistics comparing the distributions of correctly and incorrectly identified galaxies with the physical properties can be found in Table \ref{table:ks:sdss-t-eagle}. There is an obvious splitting in the distributions of M$_{\star}$ for correctly and incorrectly identified objects: the high mass merging objects are preferentially assigned the non-merger classification, while the intermediate mass non-merging objects are preferentially assigned the merger classification. Although no low mass mergers being assigned the non-merging class is reassuring as there are no low mass non-merging objects. This splitting may arise from the training sample having non-merging systems as preferentially high mass and merging objects as preferentially intermediate and low mass. A similar, but opposite, split is seen with sSFR: low sSFR mergers are identified as non-mergers, as seen in Fig. \ref{fig:sdss-t-eagle-sfr}a, while high sSFR non-mergers have a higher misclassification rate than low sSFR non-mergers, as seen in Fig. \ref{fig:sdss-t-eagle-sfr}b. This suggests that the EAGLE images for merging systems may preferentially show boosted sSFR.

\begin{table}
	\caption{KS-test statistic, D$_{N,M}$, and the critical value, Crit$_{N,M} = c(\alpha) \sqrt{\frac{n+m}{nm}}$, for the SDSS images classified by the simulation network. If D$_{N,M}$ > Crit$_{N,M}$, the null hypothesis that the two distributions are the same is rejected at level $\alpha$ = 0.05. Here, $c(\alpha)$ = 1.224 for $\alpha$ = 0.05 and $n$ and $m$ are the sizes of samples $N$ and $M$.}
	\begin{tabular}{ccccc}
		\hline
		Physical & & & & \\
		Parameter & D$_{TP,FN}$ & Crit$_{TP,FN}$ & D$_{TN,FP}$ & Crit$_{TN,FP}$ \\
		\hline
		M$_{\star}$ & 0.472 & 0.045 & 0.417 & 0.059 \\
		sSFR & 0.490 & 0.045 & 0.410 & 0.059 \\
		u-magnitude & 0.186 & 0.045 & 0.130 & 0.059 \\
		g-magnitude & 0.107 & 0.045 & 0.138 & 0.059 \\
		r-magnitude & 0.285 & 0.045& 0.268 & 0.059 \\
		i-magnitude & 0.329 & 0.045 & 0.301 & 0.059 \\
		z-magnitude & 0.379 & 0.045 & 0.334 & 0.059 \\
		\hline
	\end{tabular}
	\label{table:ks:sdss-t-eagle}
\end{table}

\begin{figure}
	\centering
	\includegraphics[width=0.5\textwidth]{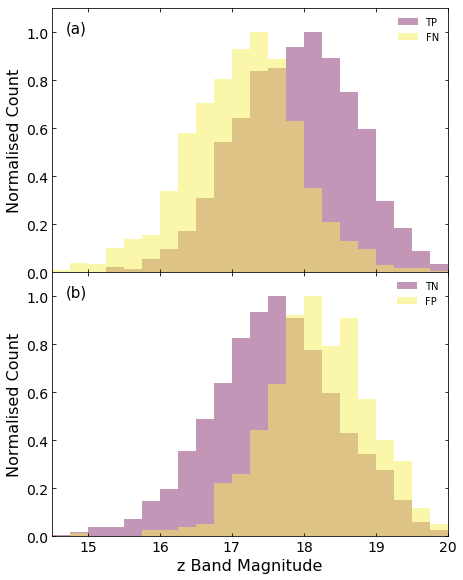}
	\caption{Distributions for the correctly (purple) and incorrectly (yellow) identified SDSS objects after being classified by the simulation network for (a) mergers and (b) non-mergers as a function of z-band magnitude. Bright mergers are preferentially classified as non-mergers while the distribution of misclassified non-mergers is skewed towards the faint end of the distribution. This trend becomes less pronounced as the bands become more blue, from z to u-band.}
	\label{fig:sdss-t-eagle-mag}
\end{figure}

The trend of the ugriz band magnitudes of the SDSS images is also interesting. As the band becomes more red, from g through to z, the distributions of correctly and incorrectly identified objects become more and more split, as can be seen by the increasing KS-test statistic in Table \ref{table:ks:sdss-t-eagle}. Thus, as the band becomes redder, more and more bright mergers are classified as non-mergers while the faint objects are correctly classified more often. Similarly, as the bands become redder, the distribution of incorrectly identified non-mergers moves to the fainter end. An example of the z-band magnitude distribution is shown in Fig. \ref{fig:sdss-t-eagle-mag}. The trend that is seen for misclassification in the merging systems is the opposite of the effect seen in the EAGLE test set when classified by the simulation network.

\begin{figure}
	\centering
	\includegraphics[width=0.5\textwidth]{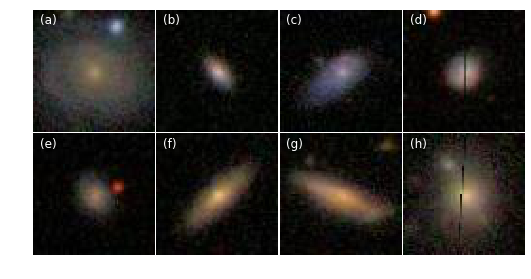}
	\caption{Examples of SDSS FP galaxies from the simulation network for (a) a galaxy with a close (in projection) companion, (b) a non-interacting, isolated galaxy, (c) a galaxy showing asymmetry or morphological disturbance and (d) a galaxy with a non-physical artifact within the image. Panels (e) to (h) show TN galaxies that are visually similar to those shown in (a) to (d).}
	\label{fig:sdss_t_eagle-fp}
\end{figure}

A subsample of FP have been visually inspected. $\approx$42\% of the FP have at least one other galaxy that lie close to the primary galaxy but are not visually interacting with the primary. These secondary galaxies are likely being identified as a merging companion to the primary or they are possibly merging systems that appear in simulations but are not identified as such in Galaxy Zoo. A further $\approx$43\% are unambiguous, non-interacting, isolated galaxies. This is possibly a result of many merging systems in the EAGLE training set visually looking like single, undisturbed galaxies. However, that does not exclude these galaxies from being true mergers as the EAGLE training set should be more complete than the SDSS images. Approximately 8\% of objects show signs of asymmetry or morphological disturbances. As with the misidentified chance projections, this may be a result of the strict selection for merging SDSS systems ignoring these galaxies but the more complete selection from EAGLE identifying these as mergers. The remaining galaxies contain a non-physical artifact, typically a single pixel width black line through the galaxy, although there are also a number of TN that also have similar artifacts, so this is unlikely to be causing the misclassification. Example FP galaxies can be found in Fig. \ref{fig:sdss_t_eagle-fp}a-d and the visually similar TN in Fig. \ref{fig:sdss_t_eagle-fp}e-h.

\begin{figure}
	\centering
	\includegraphics[width=0.5\textwidth]{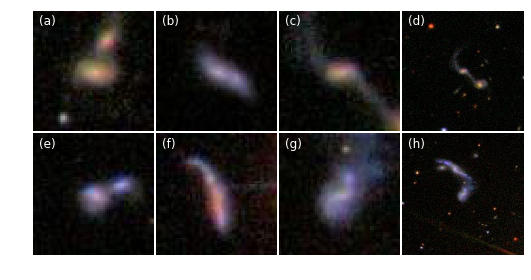}
	\caption{Examples of SDSS FN galaxies from the simulation network for (a) a galaxy with a clear merging counterpart, (b) a clearly disturbed system, (c) a galaxy whose merger companion is outside of the 64$\times$64 pixel image and (d) the larger 256$\times$256 pixel image showing the merger companion outside of panel (c). Panels (e) to (h) show TP galaxies that are visually similar to those shown in (a) to (d).}
	\label{fig:sdss_t_eagle-fn}
\end{figure}

Alongside the 526 FP, there are 1600 FN of which we visually examine a subsample. The majority of these systems ($\approx$79\%) clearly show two interacting galaxies, which may be a result of the network identifying these as chance projections. A further $\approx$7\% show clear evidence of morphological disturbances or asymmetry. Approximately 14\% of the FN have their counterpart of the edge or outside the image cut-out. This suggests, like the observation network, that a larger cut-out may help identify these objects. Examples of these objects can be found in Fig. \ref{fig:sdss_t_eagle-fn}a-d and examples of TP galaxies that look similar in Fig. \ref{fig:sdss_t_eagle-fn}e-h.

As mentioned in Sect. \ref{sec:eagle_t_sdss}, the size of the galaxies from EAGLE may be impacting the results. The re-projection of the high snapshot redshift galaxies to lower redshifts can result in the EAGLE galaxies being too small by up to a factor of two in the most extreme cases \citep{2017MNRAS.465..722F}. As the simulation network is trained on these apparently smaller galaxies, it may have difficulty in correctly identifying the larger SDSS galaxies. As before, if this limitation was reduced or removed, the result may improve.

\subsection{Differences in Network Merger Identification}
\begin{figure}
	\centering
	\includegraphics[width=0.5\textwidth]{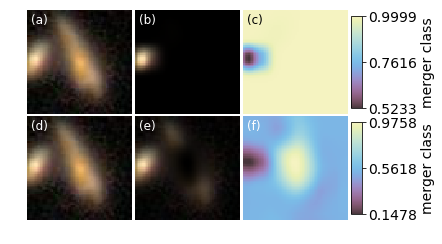}
	\caption{Heat maps to demonstrate how the observation (top row) and simulation (bottom row) networks detect an example merging SDSS system. Panels (a) and (d) show the original image of the galaxy being classified, panels (b) and (e) show the regions that most effect the merger classification and panels (c) and (f) show the heat maps where regions with darker colours have a greater affect on the classification (lower merger class output). Panels (b) and (e) are created by stretching the heat map between 0 and 1 and multiplying this with the original image.}
	\label{fig:sdss_heat}
\end{figure}

To further examine the differences in the observation and simulation networks, we examine the features that each network uses to identify a merging galaxy. A 12$\times$12 pixel area within the images that were correctly identified by both networks is made black by setting the RGB values to zero. This region is moved across the image in steps of one pixel in both x and y directions, generating 2704 images with different 12$\times$12 pixel areas masked. For each masked image, the output for the merger class is recorded. To determine which regions of the image have the greatest effect on the classification, we generate a heat map for each object where each pixel is the average of the merger class outputs of the images where that pixel is masked, as seen in the example in Fig. \ref{fig:sdss_heat}c and f.

As can be seen with an example SDSS galaxy in Fig. \ref{fig:sdss_heat}, both networks require the secondary galaxy to correctly identify the system as merging. However, as can be seen clearly in Fig \ref{fig:sdss_heat}f, the simulation network is also affected by the edges of the primary galaxy, the more diffuse regions, and uses these to help determine the classification. This is likely a result of the EAGLE merging galaxies used to train the simulation network containing systems that are longer from the merger event and so have settled somewhat as well as alignments of the EAGLE image such that the secondary galaxy is in the line of sight with the primary galaxy, requiring closer examination of the primary galaxy to find perturbations that mark them as merger. As a result, the simulation network is more sensitive to smaller changes in the profile of the more diffuse material of a galaxy and more easily fooled into providing an incorrect classification.

A similar distinction between how the observation and simulation networks function is seen when doing a similar examination with the EAGLE galaxies. The simulation network is more easily fooled by masking pixels closer into the nucleus of the primary galaxy than the observation network. However, both networks can be fooled with many galaxy images requiring the background noise to be un-masked to generate a correct classification. This indicates that the EAGLE galaxies are not a true recreation of the SDSS images, helping to partially explain the poor performance of the cross application.

\begin{figure}
	\centering
	\includegraphics[width=0.5\textwidth]{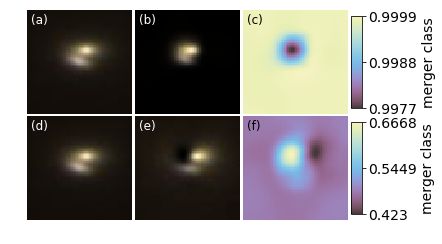}
	\caption{Heat maps to demonstrate how the observation (top row) and simulation (bottom row) networks detect an example merging EAGLE system. Panels (a) and (d) show the original image of the galaxy being classified, panels (b) and (e) show the regions that most effect the merger classification and panels (c) and (f) show the heat maps where regions with darker colours have a greater affect on the classification (lower merger class output). Panels (b) and (e) are created by stretching the heat map between 0 and 1 and multiplying this with the original image.}
	\label{fig:eagle_heat}
\end{figure}

There are also a small minority of galaxies that the observation and simulation networks need the opposite pixels to correctly identify the system as merging. For example, the simulation network may not identify the system as merging if the primary galaxy is masked but the secondary galaxy is visible while the observation network may not identify the system correctly if the secondary is masked but the primary is visible, as shown in Fig \ref{fig:eagle_heat}.

\section{Conclusions}\label{sec:conclusion}
Training and applying a CNN on SDSS images has been successful, achieving an accuracy of 91.5\%. This clearly demonstrates that CNNs can be used to reproduce visual classification. There is no clear indication of a specific type of object that is incorrectly identified from the physical or observable parameters. Training and applying a CNN on the EAGLE images was also somewhat successful, with an accuracy of 65.2\% when trained using mergers that will or have occurred within 100~Myr of the image snapshot. Using a longer time between the image snapshot and the merger reduces the accuracy of the network. This relatively lower accuracy suggests that some EAGLE mergers do not have visible merging features that can be picked up by the CNN. The incorrectly identified mergers are primarily at low simulation snapshot redshifts as well as faint apparent magnitude. The combination of real noise added to the EAGLE images and converting the absolute magnitude to apparent magnitude also reduces the effectiveness of the CNN, which demonstrates the importance of image quality (in terms of, for example, signal-to-noise and resolution) in merger identification. Within the image, chance projections result in a large number of non-merging galaxies being identified as mergers.

Examining the features of the merging galaxies that result in correct classifications, we find that the EAGLE trained network is more sensitive to features in the diffuse part of the galaxy, likely tidal structures and disturbances as well as the presence of a close companion. The SDSS trained network, however, primarily focuses only on the presence of a close companion galaxy.

The lower accuracy of the EAGLE trained network is most likely a result of the difference in the training sample. The SDSS merger sample has been selected to contain conspicuous mergers and so the features of a merger are more easily identified but will miss subtler mergers. Meanwhile, the EAGLE sample has fewer conspicuous mergers but should be more complete (including mergers with a wide range of mass ratios, gas fractions, viewing angles, environments, orbital parameters, etc.), resulting in less obvious merger features, in pixel space, that are harder for a CNN to recognise.

Passing the SDSS images through the EAGLE trained network has proven to work, although with only 64.6\% accuracy. This relatively low accuracy appears to be a result of high mass or low sSFR objects being identified as non-mergers and low mass or high sSFR objects being identified as mergers. This could suggest that simulations show evidence of high sSFR in the merging systems when this may not necessarily be true. However, the EAGLE trained network may also be identifying merging systems that the visual classification missed. The EAGLE classification will be more complete, as we know which systems are merging, and so the EAGLE trained network may be identifying these objects in the SDSS images that have been missed by the less complete, but move visually obvious, SDSS classification. The result may be a lower specificity, that is a smaller fraction of non-mergers are being correctly identified, when using the SDSS classifications as the truth when in fact the EAGLE trained network is correctly identifying merging system missed by the human visual classification. However, the relatively low recall, the fraction of mergers correctly identified, suggests that EAGLE has relatively few conspicuous mergers.

This has a tantalizing prospect for large upcoming surveys, such as LSST and \textit{Euclid}. It is possible to train a CNN with images from simulations and apply it to observations of galaxies from the real universe. Presently, the simulation trained network could be used to generate a set of galaxy merger candidates, which would need to be checked by a human expert, for use in training an observation network. However, with further refinement to the training images from simulations it is not beyond the realm of possibility to reduce the need for an observation training set and apply a simulation trained CNN directly to images from an entire survey, massively speeding up identification.

Passing the EAGLE images through the SDSS trained network was unsuccessful, with the network preferentially assigning objects to the non-merger class. This suggests that some EAGLE mergers are not representative of the SDSS selected mergers, although this appears to be primarily due to how re-projecting the galaxies to their assigned redshift has been done, so it may not be that the EAGLE mergers themselves do not look like observable mergers. The mergers in EAGLE are also less conspicuous than those in the SDSS training set so the observational network has not been trained to identify these less obvious merger events, resulting in a large number of EAGLE mergers being identified as non-mergers.

Improvements for the simulation galaxies in future work would be to increase the mass resolution, which can affect the appearance of galaxies and galaxy mergers \citep{2015MNRAS.447.3548S, 2015MNRAS.447.2753T, 2015MNRAS.452.2879T, 2016MNRAS.462.2418S}, and exactly match the stellar mass distributions with those of observations. Increasing the time resolution, for example by using the snipshots\footnote{High time resolution output from EAGLE.} instead of snapshots from EAGLE, should also provide improvement along with improving the estimates of time to or since the merger event by tracing when the central black holes merge. It would also be informative to include the effects of dust attenuation.

As has been shown in this work, chance projections are a major problem in merger classification. In future work, we could train the network to recognise chance projections better by constructing training samples of galaxies which appear close together in the sky but are actually far away from each other. Another area to improve in the future is to come up with a more refined merger classification system, rather than just a binary classification of merger versus non-mergers. For example, have separate classes of early mergers, late mergers, minor mergers, major mergers etc.

\begin{acknowledgements}
We would like to thank the anonymous referee for their thoughtful comments that have improved the quality of this paper.\\

We would like to thank S. Ellison and L. Bignone for helpful discussions that have improved this paper.\\

Funding for the SDSS and SDSS-II has been provided by the Alfred P. Sloan Foundation, the Participating Institutions, the National Science Foundation, the U.S. Department of Energy, the National Aeronautics and Space Administration, the Japanese Monbukagakusho, the Max Planck Society, and the Higher Education Funding Council for England. The SDSS Web Site is http://www.sdss.org/.\\

The SDSS is managed by the Astrophysical Research Consortium for the Participating Institutions. The Participating Institutions are the American Museum of Natural History, Astrophysical Institute Potsdam, University of Basel, University of Cambridge, Case Western Reserve University, University of Chicago, Drexel University, Fermilab, the Institute for Advanced Study, the Japan Participation Group, Johns Hopkins University, the Joint Institute for Nuclear Astrophysics, the Kavli Institute for Particle Astrophysics and Cosmology, the Korean Scientist Group, the Chinese Academy of Sciences (LAMOST), Los Alamos National Laboratory, the Max-Planck-Institute for Astronomy (MPIA), the Max-Planck-Institute for Astrophysics (MPA), New Mexico State University, Ohio State University, University of Pittsburgh, University of Portsmouth, Princeton University, the United States Naval Observatory, and the University of Washington.
\end{acknowledgements}

\bibliographystyle{aa} 
\bibliography{Paper-Real-Sim-Mergers} 

\begin{appendix}
\section{Image Colour Scaling}\label{app:linear}
As well as using a modified \citet{2004PASP..116..133L} non-linear colour normalisation to create the RGB images of the EAGLE galaxies, we also used a simple linear scaling to generate RGB images, still using gri channels for the blue, green and red colours in the images. The difference in scaling has an effect on a network's ability to accurately identify merging and non-merging galaxies.

As can be seen by comparing the results in Table \ref{table:stats} with those for the linear scaling in Table \ref{table:stats-old}, the linear scaling provides a higher accuracy and larger ROC area for the 100~Myr and 200~Myr networks, while the 300~Myr network remain approximately constant. However, the cross application with passing the linear scaled EAGLE images through the observation network shows a slight decrease in accuracy to the result presented in Table \ref{table:wrong} but the results still show a preference to assigning galaxies to the non-merger classification. Passing the SDSS images through the 100~Myr linearly scaled EAGLE trained network shows no notable difference in accuracy to the results in the main text but it does not show the tendency to assign the SDSS galaxies to the non-merger classification that the modified \citet{2004PASP..116..133L} trained network has.

Examination of the physical classifications and comparing the correctly and incorrectly identified objects, from both the self similar (EAGLE images through EAGLE trained network) and cross application, shows no qualitative differences between the modified \citet{2004PASP..116..133L} colour scaled and linearly scaled EAGLE images. Similarly, visual examination of the misclassified linearly scaled EAGLE images produced no major differences to the visual examination of the modified \citet{2004PASP..116..133L} scaled images.

\begin{table*}
	\caption{Statistics for the 100~Myr, 200~Myr and 300~Myr EAGLE trained networks at testing using EAGLE images created with a linear colour scaling. The fourth column presents the cross application of passing the linear colour scaled EAGLE images through the observation network while the fifth column presents the cross application of passing the SDSS images through the 100~Myr linearly scaled EAGLE trained network.}
	\begin{center}
		\begin{tabular}{llllcc}
		\hline
		 & & & & Linear EAGLE & SDSS images \\
		 & 100~Myr & 200~Myr & 300~Myr & images through & through linear\\
		 & & & & observation & EAGLE trained \\
		\hline
		Epoch used   &      60 &      28 &     24 &  - & - \\
		Cut threshold &   0.37 &   0.39 &  0.46 &   0.57 &   0.37 \\
		ROC area      & 0.800 & 0.745 & 0.727 & 0.515 & 0.689 \\
		Recall            & 0.667 & 0.612 & 0.652 & 0.234 & 0.676 \\
		Precision       & 0.788 & 0.715 & 0.689 & 0.494 & 0.630 \\
		Specificity      & 0.821 & 0.756 & 0.706 & 0.761 & 0.603 \\
		NPV               & 0.711 & 0.661 & 0.670 & 0.498 & 0.651 \\
		Accuracy       & 0.744 & 0.684 & 0.679 & 0.497 & 0.639 \\
		\hline
		\end{tabular}
	\label{table:stats-old}
	\end{center}
\end{table*}
\end{appendix}

 \end{document}